\documentclass[pdflatex, sn-nature]{sn-jnl}% Math and Physical Sciences Numbered Reference Style
\usepackage{graphicx}%
\usepackage{multirow}%
\usepackage{amsmath,amssymb,amsfonts}%
\usepackage{amsthm}%
\usepackage{mathrsfs}%
\usepackage{xcolor}%
\usepackage{textcomp}%
\usepackage{manyfoot}%
\usepackage{booktabs}%
\usepackage{algorithm}%
\usepackage{algorithmicx}%
\usepackage{algpseudocode}%
\usepackage{listings}%
\theoremstyle{thmstyleone}%
\theoremstyle{thmstyletwo}%

\theoremstyle{thmstylethree}%

\begin{document}

\title[SWADDL: Sampling global ocean wave state from wind history with a diffusion model]{SWADDL: Sampling global ocean wave state from wind history with a diffusion model}

%%=============================================================%%
%% GivenName	-> \fnm{Joergen W.}
%% Particle	-> \spfx{van der} -> surname prefix
%% FamilyName	-> \sur{Ploeg}
%% Suffix	-> \sfx{IV}
%% \author*[1,2]{\fnm{Joergen W.} \spfx{van der} \sur{Ploeg} 
%%  \sfx{IV}}\email{iauthor@gmail.com}
%%=============================================================%%

\author*[1]{\fnm{Jiarong} \sur{Wu}}\email{jiarong.wu@nyu.edu}

\author[2]{\fnm{Bertrand} \sur{Chapron}}\email{Bertrand.Chapron@ifremer.fr}
% \equalcont{These authors contributed equally to this work.}

\author[1]{\fnm{Laure} \sur{Zanna}}\email{laure.zanna@nyu.edu}
% \equalcont{These authors contributed equally to this work.}

\affil*[1]{\orgdiv{Courant Institute School of Mathematics, Computing and Data Science}, \orgname{New York University}, \orgaddress{\street{251 Mercer Street}, \city{New York}, \postcode{10012}, \state{NY}, \country{USA}}}

\affil[2]{\orgdiv{Laboratoire d’Océanographie Physique et Spatiale (LOPS)}, \orgname{Ifremer}, \orgaddress{\street{1625 route de Sainte-Anne}, \city{Plouzané}, \postcode{29280}, \state{Bretagne}, \country{France}}}

%%==================================%%
%% Sample for unstructured abstract %%
%%==================================%%

\abstract{
Predicting ocean wave state (sea state) is essential for maritime operations and coupled Earth system modeling, yet spectral wave models remain too computationally expensive for many applications, including online coupling to climate models and ensemble-based probabilistic prediction. Existing AI-based wave models are largely deterministic and limited to bulk wave state variables such as significant wave height. Here we present SWADDL (Sampling WAve state Distribution using Diffusion modeL), a generative model that samples global sea state conditioned on 10 days of global wind forcing, without autoregressive time-stepping. Beyond bulk variables, SWADDL estimates wave-partition variables and derived quantities such as Stokes drift and mean square slope. Trained on a 30-year global WAVEWATCH-III hindcast, SWADDL runs roughly 20 times faster than the spectral model while delivering skillful predictions and calibrated ensemble spread for bulk variables. Over six test years, SWADDL reproduces the large-scale patterns and seasonal cycle of the wave climate, and a residual variant that samples departures from a deterministic prediction further reduces regional climatological biases and achieves the lowest probabilistic error. Our results suggest that diffusion-based sea state sampling offers a promising path toward probabilistic wave forecasting and efficient coupling of sea state information into Earth system models.

% We further demonstrate model performance under extreme wave height conditions using [Hurricane Ivan?] as a benchmark. 
% Comparison with NDBC buoys?
}

\keywords{ocean waves, wave prediction, diffusion model, deep learning}

%%\pacs[JEL Classification]{D8, H51}

%%\pacs[MSC Classification]{35A01, 65L10, 65L12, 65L20, 65L70}

\maketitle

\section{Introduction}\label{sec1}
The general condition of ocean surface waves, also termed sea state, is characterized by the directional wave spectrum or, often more concisely, by spectrally averaged quantities including wave height, period, and direction, etc. Predicting sea state is essential for operational purposes, as wave conditions affect many aspects of maritime and coastal activities. It is also important for coupled earth system modeling, as surface waves can modulate exchanges between atmosphere and ocean on both weather and climate scales \citep{janssen_interaction_2004, cavaleri_wind_2012}. Such effects are reflected in various sea-state-dependent parameterizations of processes in the air-sea transition zone, such as Langmuir turbulence and breaking-enhanced mixing in the upper ocean \cite{cavaleri_wind_2012}.
% the drag coefficient for marine atmospheric boundary layer and 

Ocean surface waves obtain their energy from wind forcing, so ultimately, sea state prediction can be viewed, to first order,  as a forced problem. 
%However, different 
Resulting wave systems can then exhibit different persistent spatiotemporal scales. Low-to-moderate winds generate local wave systems with limited energy and relatively short peak wavelengths which 
%and can 
decay quickly once the wind forcing ceases. On a global scale, the longest-lived wave systems are those forced by intense meteorological events, e.g., tropical and extra-tropical storms, and can travel across ocean basins, becoming known as swells \cite{snodgrass_propagation_1966}, radiating a large amount of momentum and energy across ocean basins. Besides the use of self-similarity laws under stationary wind \citep{kitaigorodskii_application_1962, dulov_fetch-_2020} or oversimplified moving storm conditions \citep{yurovskaya_self-similar_2023}, this complex dependence of sea state on spatially and temporally varying (non-local) wind history is hard to describe with closed-form equations. Therefore, most wave models solve the evolution of the sea state using a discrete spatial and temporal grid, with local and concurrent wind forcing as the source term. In this sense, the task is closer to nowcasting than forecasting, though we use the two terms interchangeably throughout this paper.

There are varying levels of complexity in sea state description, ranging from a full 2D spectrum to wave partitions to averaged bulk parameters. Operational numerical wave models (e.g., NOAA's WAVEWATCH-III \cite{the_wavewatch_iii_development_group_ww3dg_wave_2016}) solve for the full 2D energy spectrum, %with the main advantage that there exists a closed-form equation, 
following the spectral density evolution of the wave action balance equation
under a source function. The source function is written as a sum of different terms,  parameterizing  the wind growth, the wave-breaking dissipation and the energy transfer associated to non-linear wave-wave interactions.
%(although several processes require parameterization).
However, there are non-negligible downsides: the discretization in spectral space results in a daunting $\mathcal{O}(100)$ dimension state vector for each spatial grid and significant computational cost, prohibiting the deployment of spectral wave models in many other applications, such as coupled climate simulations \cite{hell_particle--cell_2025}. Moreover, there are still very limited observations of the full 2D wave spectrum, so model verification %still 
typically relies on bulk variables such as significant wave height.
At intermediate levels of complexity, reduced-order models limit themselves to the mean of each wave system (partition), while still preserving some level of multimodal sea-state description. These methods are based on parametric equations and often require complex geometric algorithms such as ray tracing \cite{kudryavtsev_2d_2021}, and are not mature enough for global-scale applications. The most simplified description is the averaged bulk (mean) wave variables. For traditional numerical methods, this is not yet feasible, since the governing equation is unknown.
% Although not previously adopted as prognostic variables in numerical models because the governing equation is unknown, this mean description is what most emerging deep learning models are based on, using either the output of numerical models or reanalysis as training data. This is because the paradigm of these computer vision based models works on pixelated images and the full 2D wave spectrum has too many degrees of freedom and may require convolution in the spectral space. 
% State-of-the-art numerical wave models (e.g., WAVEWATCH-III) are widely used in operations and can provide skillful forecasts. However, they are computationally expensive, which severely limits their deployment in coupled ocean-atmosphere models \cite{hell_particle--cell_2025} and hinders the implementation and testing of sea-state-dependent parameterizations. 

% Should we include more citations?
% Pangu-Weather (Bi et al., 2023, Nature), Huawei
% GraphCast (Lam et al., 2023, Science), open-sourced by Google DeepMind
% FourCastNet (Pathak et al., 2022 and Bonev et al., 2025)), Nvedia
% Fuxi a series of papers

The emergence of AI-based numerical weather forecasting and earth system modeling \cite{kochkov_neural_2024, bodnar_foundation_2025, wang_physics-guided_2024, wang_data-driven_2025, zhang_ocean_2025, hahner_representing_2026} has applied deep learning methods that can extract complex statistical relationships from large datasets and capture long-range spatiotemporal dependencies.
This has important implications for sea-state estimation, as it enables more flexible model design choices beyond the established but costly, high-dimensional spectral paradigm. In fact, contrary to conventional numerical methods, recent works using deep learning for ocean wave forecasting \cite{kochkov_neural_2024, bodnar_foundation_2025, wang_physics-guided_2024, wang_data-driven_2025, zhang_ocean_2025, hahner_representing_2026} often use bulk variables, as this allows the direct adoption of frameworks already developed for meteorological variables in weather forecasting. Alongside the most common autoregressive rollout framework \citep{bodnar_foundation_2025, wang_data-driven_2025, zhang_ocean_2025, hahner_representing_2026} with additional forcing fields at each time step, there has been work that shows success in using only a long wind history to directly infer wave height \citep{wang_physics-guided_2024}. 

\begin{figure}[!ht]
\centering
\includegraphics[width=1\textwidth]{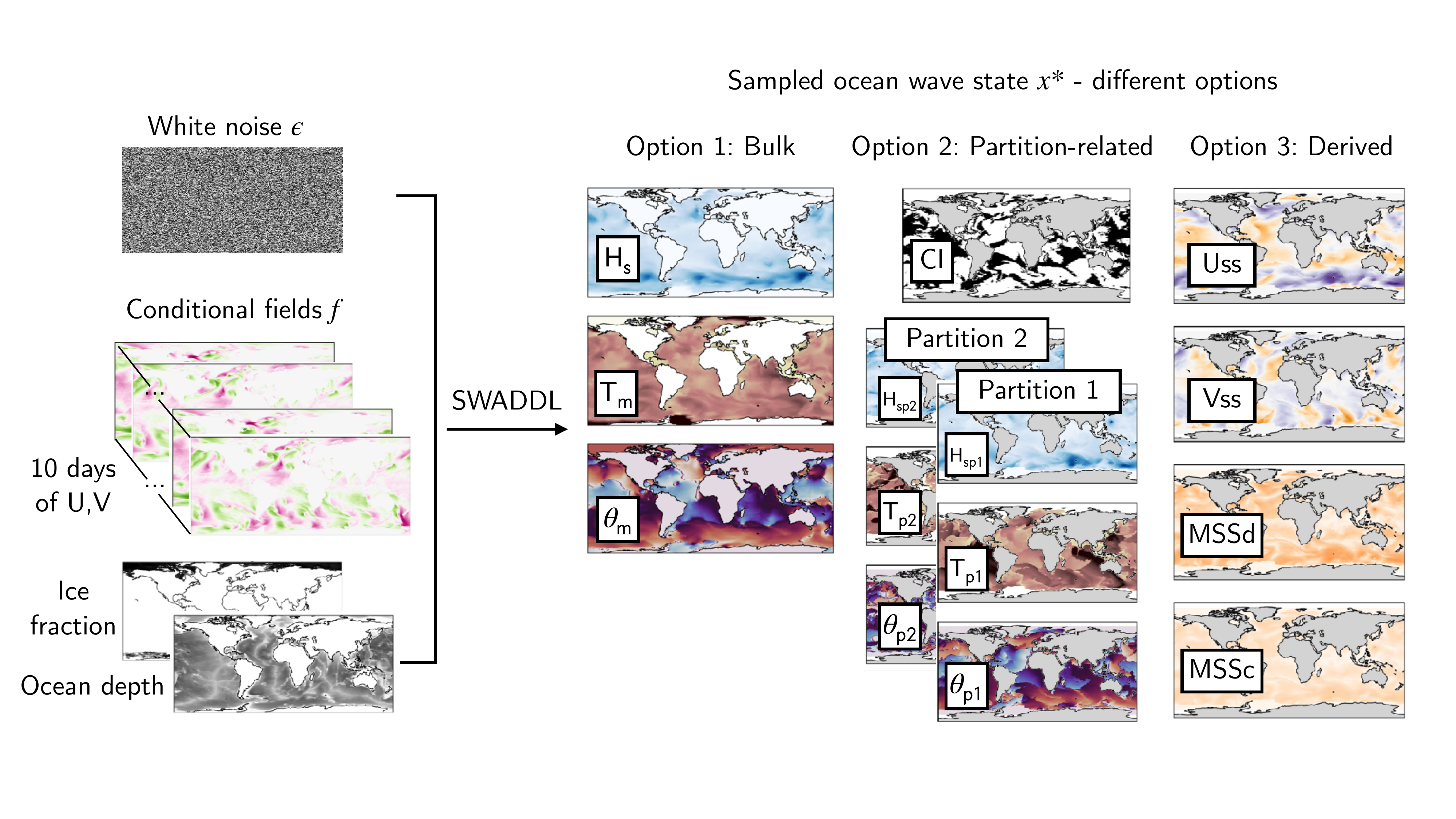}
\hspace{-1cm}
\caption{Denoising diffusion model-based sea state sampler SWADDL, conditioned on global wind history and auxiliary fields (ice fraction and ocean depth). It enables flexible sampling of an array of sea state variables beyond the basic bulk variables. See Section \ref{subsec:variables} for the definitions of bulk, partition-related, and derived variables. Each option is predicted using a seperately trained diffusion model. The denoiser network follows a U-Net like structure, see Section \ref{subsec:DDPM} for detail.}\label{fig:diagram}
\end{figure}

The objective of this work is to leverage recent advances in deep learning to provide efficient sea state estimation at a global scale. As mentioned, this will not only benefit operational forecasting but also applications such as coupling waves to climate models, where sea state information is needed but currently too costly to access online. Beyond bulk wave variables, we aim to provide estimates of an array of additional variables that characterize different aspects of the sea state, including partition-related and derived variables (see their definitions in Section \ref{subsec:variables}) as illustrated in Figure \ref{fig:diagram}. The partition-related variables provide a more refined description of the sea state, especially when multiple collocated wave systems are present at a given location. The derived variables are those that enter various sea-state-dependent parameterizations when coupling waves to other Earth system models, and we use Stokes drift and mean square slope as demonstrations. 
We train our model on a multi-decade hindcast \cite{rascle_global_2013, alday_global_2021} performed with WAVEWATCH-III (a spectral wave model), in which all these variables are available as model outputs.
% The goal is to replace the expensive spectral wave models, not to do better.

In the realm of numerical weather forecasting, it is generally recognized that ensemble-based probabilistic forecasting methods are superior to deterministic ones \cite{leutbecher_ensemble_2008}. In this spirit, probabilistic AI weather models are actively being developed \cite{kochkov_neural_2024, price_probabilistic_2025, lang_aifs-crps_2026}. They can provide useful information about forecast uncertainties and are less prone to over-smoothing in small-scale structures that accompany many deterministic mean-square-error-based models. Among probabilistic approaches, score-based models \cite{song_score-based_2020, ho_denoising_2020}, especially diffusion models designed to sample from conditional distributions, have gained traction in applications beyond ensemble weather forecasting, including data assimilation and climate state sampling \cite{price_probabilistic_2025, rozet_score-based_2023, brenowitz_climate_2025}. 
By contrast, probabilistic approaches to wave forecasting remain largely unexplored. This is likely due in part to the perception that waves are predominantly wind-driven, so that wind uncertainty alone governs wave uncertainty, compounded by the prohibitive computational cost of running spectral wave models in an ensemble setting. The absence of probabilistic sea-state prediction consequently limits the application of state-of-the-art data assimilation techniques in wave modeling \cite{houghton_operational_2022}.
% Mention CRPS? Most of them use Continuous Ranked Probability Score (CRPS). 
% Next to providing useful information about forecast uncertain-ties, these models have more stable statistics than the deterministically trained models, as shown by for example spectra of forecast fields, and do not smooth out small-scale structures with forecast lead time.
% Diffusion model has, to our knowledge, not been applied to sea state estimation in climate context, and this offers an exciting interdisciplinary opportunity to tackle the long-standing problem.

In this work, we propose a sea state sampling approach based on the denoising diffusion model \cite{ho_denoising_2020}. By conditioning on a relatively long history of global wind field, the diffusion process directly samples the complex data distribution without resorting to autoregressive time-stepping. The length of wind history is chosen based on an approximate time scale of swell propagation across the ocean basin \cite{snodgrass_propagation_1966}. 
This conditional diffusion framework provides flexibility to sample sea state variables of interest beyond the bulk variables, as illustrated in Figure \ref{fig:diagram}. We demonstrate that the proposed sea state sampler SWADDL (Sampling WAve state Distribution using Diffusion modeL) achieves significant acceleration over numerical spectral models and provides relatively skillful predictions. We discuss the varying skill and calibration of the ensemble across different sea state variables.  
% GenCast is implemented as a conditional diffusion model, a generative ML method that can model the probability distribution of complex data and generate new samples. 

\section{Results}\label{sec2}
\begin{figure}[!ht]
    \centering
    \includegraphics[width=1\linewidth]{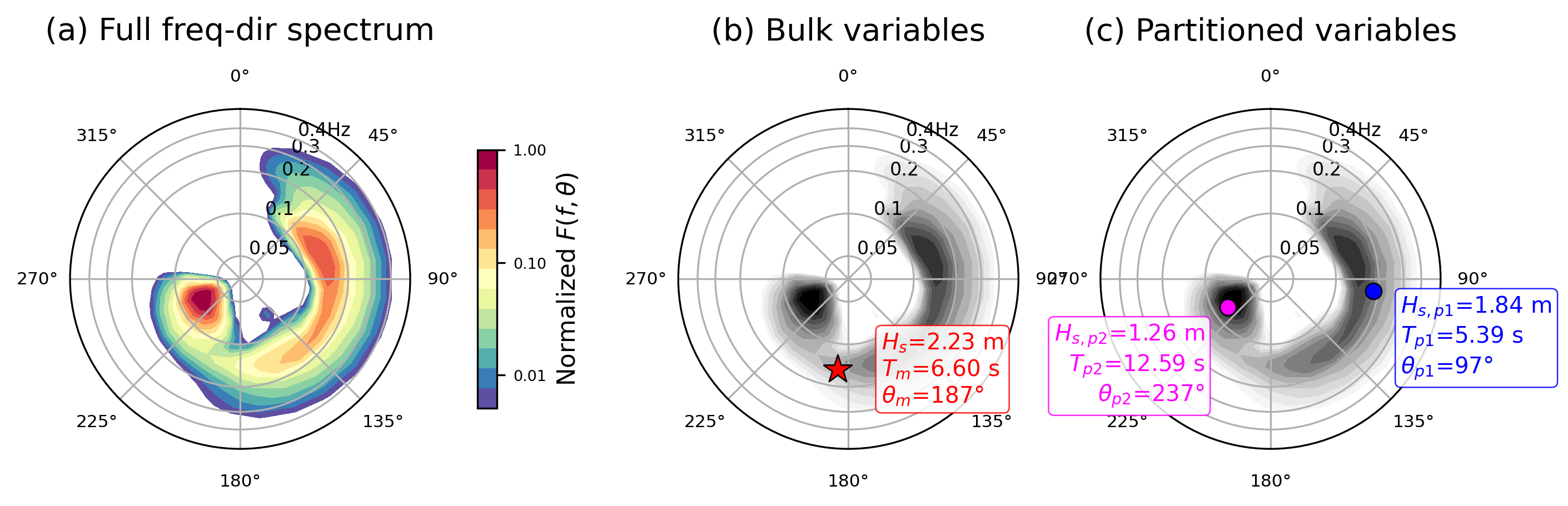}
    \caption{Illustration of bulk (b) and partitioned (c) variables, based on an example frequency-direction spectrum (a).}
    \label{fig:var_visual}
\end{figure}

The sea state at a given location is a random superposition of waves with different frequencies and directions, fully described by the frequency--direction spectrum $F(f,\theta)$. Each variable predicted here is an integral of this spectrum that summarizes
a distinct physical aspect of the wave field (Fig.~\ref{fig:var_visual}; mathematical definitions in Section~\ref{subsec:variables}). Among the bulk variables, $H_s$ measures the total wave energy, $T_m$ is an energy-weighted mean period, long for swell and short for young wind sea, and $\theta_m$ is the dominant direction of wave energy. 

When local wind sea coexists with swell from distant storms, the spectrum is multi-modal, and bulk averages can describe a period or direction lying between two distinct wave systems (Fig.~\ref{fig:var_visual}b). Partition variables are a better representation in this situation by reporting $H_{sp}$, $T_p$, and $\theta_p$ for each of the two most energetic wave systems (Fig.~\ref{fig:var_visual}c). 
The crossing sea index (CI) flags locations where two wave systems coexist with comparable energy levels propagating in distinctly different directions \cite{hanson_automated_2001, cavaleri_rogue_2012}, a condition with implications for air-sea interfacial processes and for the occurrence of extreme wave conditions such as rogue waves \cite{cavaleri_windwave_2020}.

% from ERA5 reanalysis

There exist many additional derived variables from the full spectrum, and here we discuss two of them: surface Stokes drift and mean square slope. Surface Stokes drift is the net transport of water parcels induced by wave orbital motion. It is the dominant contribution to wind-induced surface drift \cite{rascle_global_2013} and drives Langmuir circulation in the upper ocean \cite{cavaleri_wind_2012}. Mean square slope (MSS) is the variance of the surface slope and measures surface roughness. It is linked to wave breaking and air--sea gas exchange \cite{garbe_transfer_2014}, and to the scattering of sunlight and radar signals used in remote sensing \cite{munk_inconvenient_2009}.
% and affects the retrieval of Synthetic Aperture Radar (SAR) imaging signals. 

In the following sections, we discuss results for each group of sea state variables. The verification metrics are defined in Section \ref{sec:metrics}: the root-mean-square error (RMSE) and anomaly correlation coefficient (ACC) evaluate the skill of the ensemble mean; the continuous ranked probability score (CRPS) is a proper score used to evaluate the quality of ensemble forecast; and the spread-skill ratio (SSR) quantifies how well the ensemble is calibrated. We use years 1993 to 2017 of the hindcast for training, year 2019 for validation, and years 2020 to 2025 for testing. The statistics shown in figures and tables throughout the paper are averaged between 2020 to 2025, if not otherwise specified.  
\subsection{Bulk sea state variables}
\begin{figure}[!ht]
\centering
\includegraphics[width=1\textwidth]{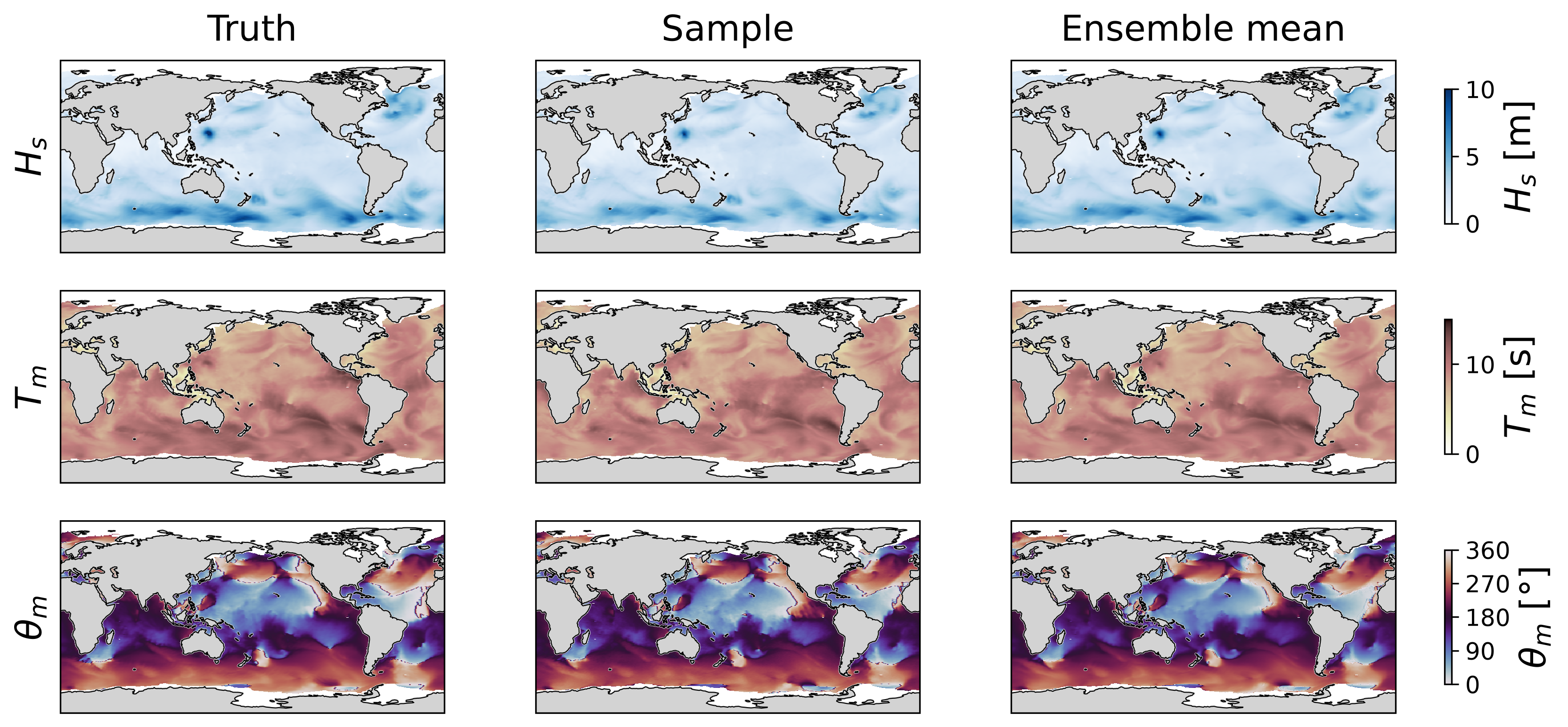}
\caption{Example snapshot of sampled bulk sea state variables and their ensemble mean, compared with the verification target (truth) for 2022-04-14-00UTC. Super Typhoon Malakas east of the Philippines, near its peak intensity.}\label{fig:mean_var}
\end{figure}
In Figure \ref{fig:mean_var}, we plot an example of bulk sea state variables for 2022-04-14, showing one individual sample and the ensemble mean, in comparison to the hindcast target. SWADDL captures all three variables relatively well, even for the extreme wave height generated by Super Typhoon Malakas. We observe similar performance for other tropical cyclones in the test set, though we do not show them here for brevity.

Table \ref{tab:mean_var} lists the verification metrics for SWADDL alongside reference values from an autoregressive deep learning model in the literature \cite{zhang_ocean_2025}. The RMSE of SWADDL is reasonably low, comparable to typical discrepancies between observations and wave model hindcasts \cite{rascle_global_2013, wang_data-driven_2025}. The ACC is highest for $H_s$ and lower for $T_m$ and $\theta_m$, indicating that significant wave height is the easiest variable for the model to predict. When compared to the autoregressive model of \cite{zhang_ocean_2025}, SWADDL achieves comparable RMSE to their 360-hour forecast but larger RMSE than their shorter-lead-time forecasts. This is expected, as SWADDL generates each snapshot independently from wind forcing alone, without information from prior wave states.
% \textcolor{blue}{Question of predictability.}

\begin{table}[h]
\caption{Verification metrics for the bulk (mean) variables, evaluated on 2020--2025 every 5 days (00~UTC, 439 dates) with 10 ensemble members. RMSE and ACC are those of the ensemble mean, and ACC is computed from anomalies relative to the monthly climatology of the training period (1993–2017); all metrics are latitude weighted over ice-free ocean; $\pm$ is the standard deviation of the daily RMSE.} \label{tab:mean_var}%
\begin{tabular}{@{}lllllll@{}}
\toprule
      & RMSE  & ACC & CRPS & SSR & RMSE ref \footnote{1} \\
\midrule
$H_s$ [m]      & 0.24 $\pm$ 0.04 & 0.96 & 0.11 & 0.93 & [0.09, 0.22]  \\
$T_m$ [s]      & 0.71 $\pm$ 0.13 & 0.83 & 0.34 & 0.93 & [0.25, 0.70]  \\
$\theta_m$ [$^{\circ}$] & 20.25 $\pm$ 3.14 & 0.78 & 7.02 & 0.93 & [17.5, 39.0]  \\
\botrule
\end{tabular}
\footnotetext[1]{As a function of lead time from 0 to 360 hours (see Figure 2 of \cite{zhang_ocean_2025}). Note that ACC listed here cannot be directly compared to the same reference since it is computed against monthly instead of long-term climatology.}
\end{table}

\begin{figure}[h]
\centering
\includegraphics[width=1\textwidth]{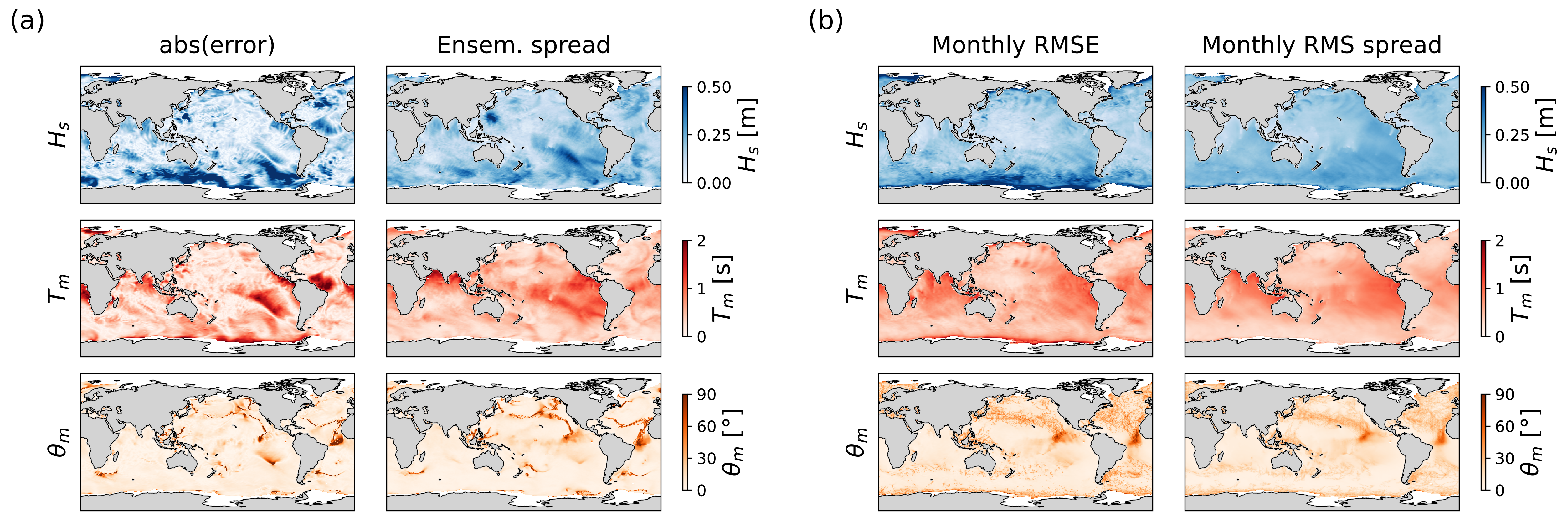}
\caption{(a) Absolute error of the ensemble mean and ensemble spread for the bulk variables, shown for 2022-04-14-00UTC. (b) Monthly RMSE and monthly RMS spread for April of 2022. The ratio of the two corresponds to SSR.} \label{fig:mean_err_std}
\end{figure}

To examine whether the probabilistic model is well calibrated, we compare maps of errors and ensemble spread, and compute the SSR for each bulk variable. Figure \ref{fig:mean_err_std}a shows the absolute error of the ensemble mean and the ensemble spread on 2022-04-14. Since the error on a single date is one realization of a random variable whose expected magnitude is set by the spread, it is spatially noisier than the spread, but large errors concentrate where the spread is large, for example around Typhoon Malakas and along swell fronts. Over a month (Figure \ref{fig:mean_err_std}b), the RMSE of the ensemble mean and the RMS ensemble spread, which should agree for a calibrated ensemble, show similar spatial patterns and magnitudes. The correspondence is particularly clear for the direction variable, suggesting that the model has learned to identify regions where multiple wave systems intersect, leading to greater directional uncertainty. The main exception is the Southern Ocean, especially near the sea-ice edge, where the RMSE of $H_s$ exceeds the spread because SWADDL underestimates $H_s$ there (see Section~\ref{subsec:climatology}); this bias is not represented by the ensemble spread. Globally, the SSR is 0.93 for all three variables (Table \ref{tab:mean_var}), indicating a nearly calibrated, slightly under-dispersed ensemble.

% Fully developed waves.
% Do we achieve comparable RMSE errors compared to deterministic forecasting? 
% Show a map with wave age.

\begin{figure}[h]
\centering
\includegraphics[width=0.7\textwidth]{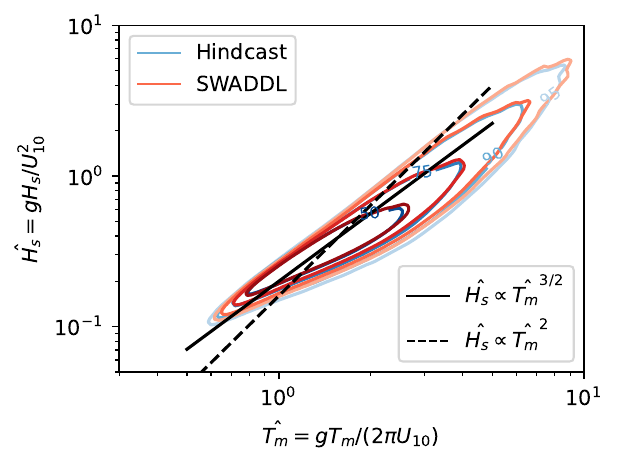}
\caption{Global distribution of non-dimensional significant wave height v.s. non-dimensional mean wave period for both the hindcast data (target) and SWADDL samples. Solid line: Toba's law $\hat{H_s} \propto \hat{T_m}^{3/2}$; dashed line: saturated waves $\hat{H_s} \propto \hat{T_m}^{2}$. Density contour lines are plotted corresponding to the indicated percentile levels in the hindcast data. }\label{fig:hs_vs_tp}
\end{figure}

A further sanity check examines one aspect of the sea state bulk variable distribution: the physical consistency of the predicted significant wave height $H_s$ and mean wave period $T_m$. When normalized by 10-meter wind speed $U_{10}$ and gravitational acceleration $g$, 
\begin{equation}
    \hat{H_s} = gH_s/U_{10}^2, \; \hat{T_m} = gT_m/(2\pi U_{10}),
\end{equation}
the joint distribution of non-dimensional parameters $\hat{H_s}$ and $\hat{T_m}$ is constrained by wind-wave growth physics, and several scaling relations have been suggested \cite{badulin_physical_2014}. This joint distribution is plotted in Figure \ref{fig:hs_vs_tp}, together with the famous Toba's 3/2 power law for growing wind sea \cite{toba_local_1972}, and the square power law for fully developed waves with saturated mean slope \cite{badulin_physical_2014}. As can be seen from the joint distribution, the global wave data are more complicated than these simple reductive relations, especially with the prevalence of swells causing data to fall off these lines (towards larger $\hat{T_m}$). Figure \ref{fig:hs_vs_tp} suggests that this non-trivial joint distribution in the hindcast data is well preserved by SWADDL, with almost overlapping density contours. 

\subsection{Partition-related variables}
A key question is whether the diffusion model can predict co-existing wave systems at a given location, i.e., multi-modality in the underlying wave spectrum.
First, we discuss the probabilistic prediction of a binary label, the crossing sea index, for the occurrence of misaligned co-existing wave systems. Then we examine the performance for predicting the mean variables of the two most energetic wave systems.
\subsubsection{Crossing sea index}
Both the prediction $x_{i,j,k}^m$ and the target $y_{i,j,k}$ for CI are binary, and the ensemble mean $\bar{x}_{i,j,k} \in [0,1]$ can be interpreted as the predicted probability of crossing sea occurrence. Crossing sea occurs in about 43\% of the grid points in the hindcast dataset, and is most common in the Tropics. 

\begin{figure}[h]
\centering
\includegraphics[width=1\textwidth]{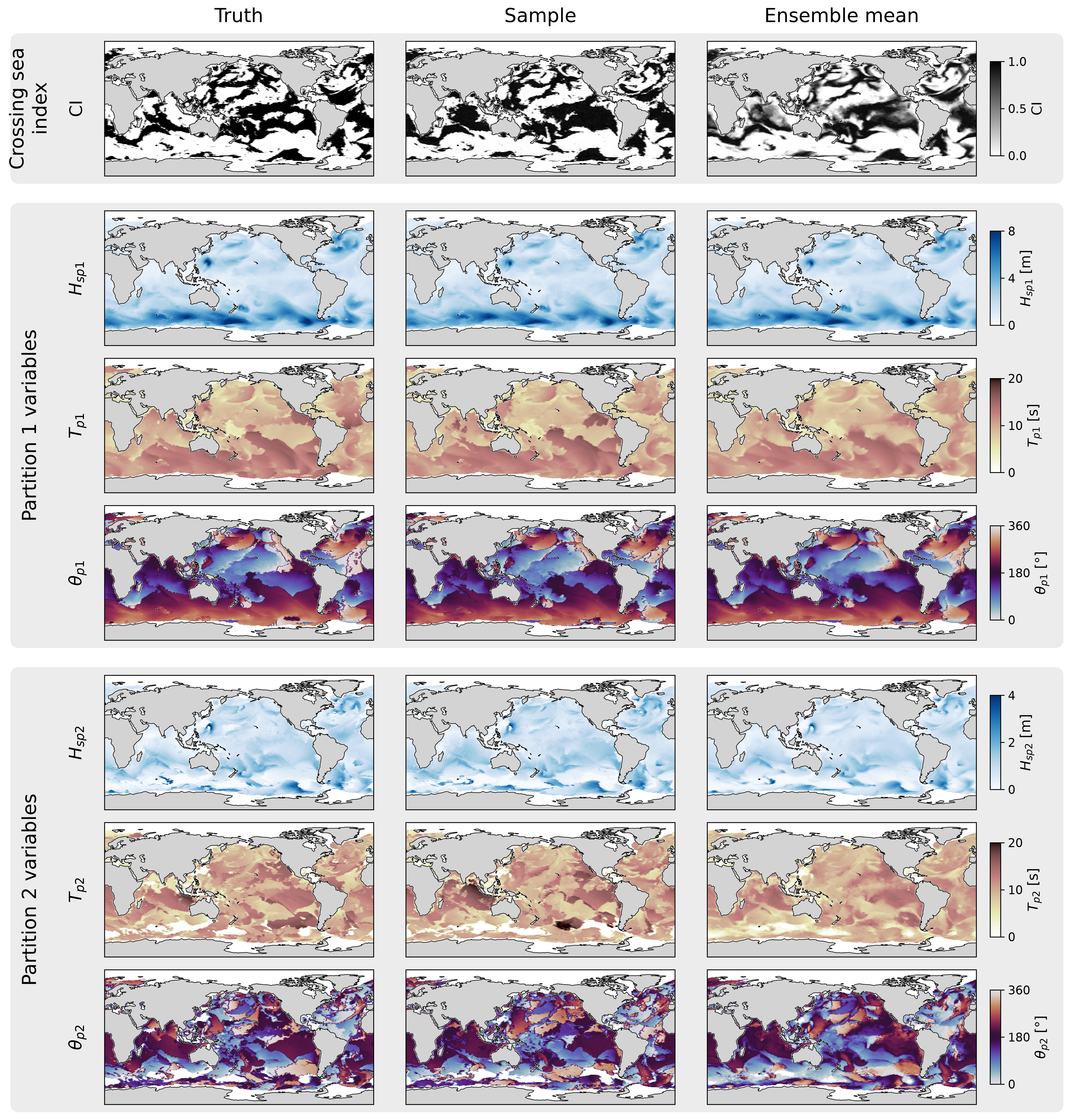}
\caption{Example snapshot of sampled partition sea state variables (crossing sea index, and the variables for the two most energetic partitions) and their ensemble mean, compared with the verification target (truth) for 2022-04-14-00UTC.\label{fig:partition}}
\end{figure}

As shown in Figure \ref{fig:partition}, SWADDL provides a reasonable estimate of this probability. The model shows significant skill relative to monthly climatology with a Brier Skill Score (see definition in Section \ref{sec:metrics}) of approximately 0.4. Although CI is the most basic indicator of multi-model sea states, the model's ability to predict it suggests that it has learned to distinguish regions and conditions in which distinct wave systems coexist, which is encouraging for the partition-level predictions discussed next.
% \textcolor{blue}{(monthly breakdown?)}

\subsubsection{Mean variables of partitions}
The variables characterizing the most energetic system (partition 1) and the second-most energetic system (partition 2) are plotted in Figures \ref{fig:partition} (see definition in Section \ref{subsection:partition_var}). Clearly, the partitioned fields are much less smooth spatially than the global bulk variables and considerably more difficult to predict. 

Partition 1 exhibits longer mean wave periods characteristic of swell-dominated sea states, reflecting well-defined spectral peaks that are obscured by broadband averaging in bulk variables. As shown in Figure \ref{fig:partition}, the diffusion model captures these long-period swell components in individual ensemble members, though the signal is attenuated when computing the ensemble mean. Certain regions in Figure \ref{fig:partition} are empty, corresponding to areas dominated by a single wave system. These areas are well represented by the diffusion model, as it has learned a representation of single vs. multi-modal sea state, similar to that necessary for predicting crossing-sea index.

The verification metrics for partitions 1 and 2 are listed in Table \ref{tab:partition}. As expected, the errors are larger than for the bulk variables. In addition, the ensemble exhibits under-dispersion, with SSR values consistently smaller than 1. To our knowledge, this is the first attempt to provide partition-level prediction at a global scale, and no baseline from previous work exists for direct comparison. The larger errors likely reflect both the intrinsic difficulty of predicting partition-level fields, and limitations in the current model architecture. An important direction for future work is the development of joint evaluation metrics that assess the two partitions together rather than independently, as the labeling of partition 1 versus 2 is itself subject to ambiguity when multiple wave systems carry comparable energy.

\begin{table}[!ht]
\caption{Verification metrics for the partitioned variables, evaluated on 2020--2025 every 5 days (00~UTC, 439 dates) with 10 ensemble members, as in Table~\ref{tab:mean_var}.} \label{tab:partition}
\begin{tabular*}{\textwidth}{@{\extracolsep\fill}lcccccc}
\toprule%
& \multicolumn{3}{@{}c@{}}{Partition 1} & \multicolumn{3}{@{}c@{}}{Partition 2} \\\cmidrule{2-4}\cmidrule{5-7}%
Variables & $H_{sp1}$ [m] & $T_{p1}$ [s] & $\theta_{p1}$ [$^{\circ}$] & $H_{sp2}$ [m] & $T_{p2}$  [s] & $\theta_{p2}$ [$^{\circ}$]  \\
\midrule
RMSE & 0.25 $\pm$ 0.03 & 1.48 $\pm$ 0.23 & 29.91 $\pm$ 4.30 & 0.28 $\pm$ 0.03 & 2.94 $\pm$ 0.29 & 57.08 $\pm$ 3.73 \\
ACC  & 0.97 & 0.77 & 0.73 & 0.82 & 0.58 & 0.54 \\
CRPS & 0.12 & 0.60 & 10.07 & 0.13 & 1.32 & 24.84 \\
SSR  & 0.90 & 0.86 & 0.90 & 0.85 & 0.85 & 0.91 \\
\botrule
\end{tabular*}
\end{table}

\subsection{Derived variables}
\begin{figure}[!ht]
\centering
\includegraphics[width=1\textwidth]{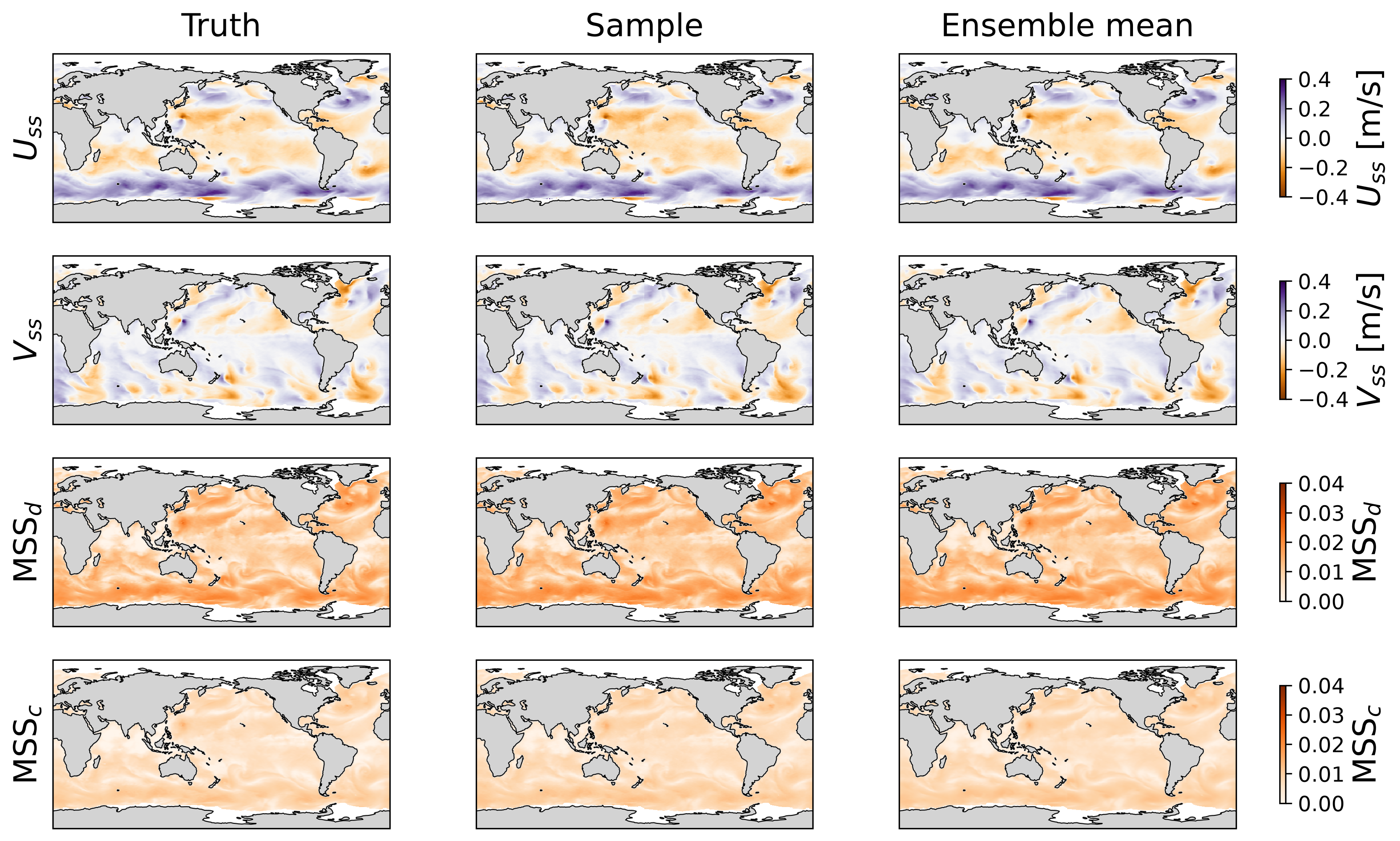}
\caption{An example of sampled surface Stokes drift and mean square slope and their ensemble mean, as compared to the verification targets (truth) for 2022-04-14-00UTC.}\label{fig:derived_var}
\end{figure}

Figure \ref{fig:derived_var} shows that both the zonal and meridional components of Stokes drift, as well as the down-wave ($\text{MSS}_d$) and cross-wave ($\text{MSS}_c$) components of MSS are well sampled by the diffusion model, with individual ensemble members closely resembling the ensemble mean, indicating small ensemble spreads for these variables.

We further examine the effect of including wind history for these derived variables. Since surface Stokes drift and MSS are higher-order moments of the spectrum, they are expected to be determined more dominantly by the shorter waves than lower-order moments, such as significant wave height. Shorter waves are more responsive to local wind forcing; consequently, we expect that conditioning on a long wind history is less important for these higher-order statistics. As shown in Table \ref{tab:derived_var}, including the wind history improves the RMSE of surface Stokes drift (third moment) but degrades the RMSE of MSS (fourth moment). This indicates that the benefit of including wind history diminishes as the target shifts towards higher spectral moments. The ensemble of MSS also appears to be under-dispersed both with and without wind history, suggesting that, in practice, a deterministic deep learning framework might suffice for predicting this particular target.

\begin{table}[h]
\caption{Verification metrics for the derived variables without and with wind history, evaluated on year 2023 with every 5 day sampling.} \label{tab:derived_var}%
\begin{tabular*}{\textwidth}{@{\extracolsep\fill}lcccccccc}
\toprule
& \multicolumn{4}{@{}c@{}}{Without wind history} & \multicolumn{4}{@{}c@{}}{With wind history}
\\\cmidrule{2-5}\cmidrule{6-9}%
      & RMSE & ACC & CRPS & SSR & RMSE & ACC & CRPS & SSR \\
\midrule
$U_{ss}$ [$10^{-2}$ m/s] & $1.05 \pm 0.05$ & 0.985 & 0.55 & 1.06 & $0.76 \pm 0.05$ & 0.992 & 0.40 & 1.18 \\
$V_{ss}$ [$10^{-2}$ m/s] & $0.96 \pm 0.06$ & 0.987 & 0.50 & 1.03 & $0.75 \pm 0.04$ & 0.992 & 0.40 & 1.08 \\
$\text{MSS}_d$ [$10^{-3}$] & $0.92 \pm 0.13$ & 0.968 & 0.49 & 0.69 & $1.29 \pm 0.21$ & 0.934 & 0.89 & 0.42 \\
$\text{MSS}_c$ [$10^{-3}$] & $0.74 \pm 0.08$ & 0.932 & 0.39 & 0.69 & $0.86 \pm 0.12$ & 0.907 & 0.52 & 0.49 \\
\botrule
\end{tabular*}
\end{table}

\subsection{Wave climatology and residual DDPM model}\label{subsec:climatology}

The examples in the previous sections are single snapshots and offer qualitative evidence of forecast quality, while the verification scores in Tables~\ref{tab:mean_var} and \ref{tab:partition} are global long-term means and can mask regional or seasonal errors. In this section, we evaluate the models over the full test period (2020--2025, every 5 days, 439 dates) as maps of time-mean skill and as regional seasonal cycles. We focus on the time mean and seasonal variability rather than multi-year trends, because global trends are small \cite{young2011} and difficult to detect over the test period. Note that the seasonal and interannual variability and the trends in the hindcast originate from the ERA5 wind forcing, so these tests assess whether the learned wind-to-wave mapping transfers the variability of the wind to the wave field. This is the property that matters for coupling to climate models, in which the wave sampler should respond to changes in the simulated winds. We compare three models that share the same U-Net architecture and training data (Section~\ref{subsec:DDPM}): the stand-alone DDPM used in the previous sections, the deterministic model, and the residual diffusion model (res-DDPM), which samples the departure from the frozen deterministic prediction. The res-DDPM is trained from scratch for 10 epochs with the same noise schedule and optimizer settings as the stand-alone DDPM. 

\begin{figure}[!ht]
    \centering
    \includegraphics[width=1\linewidth]{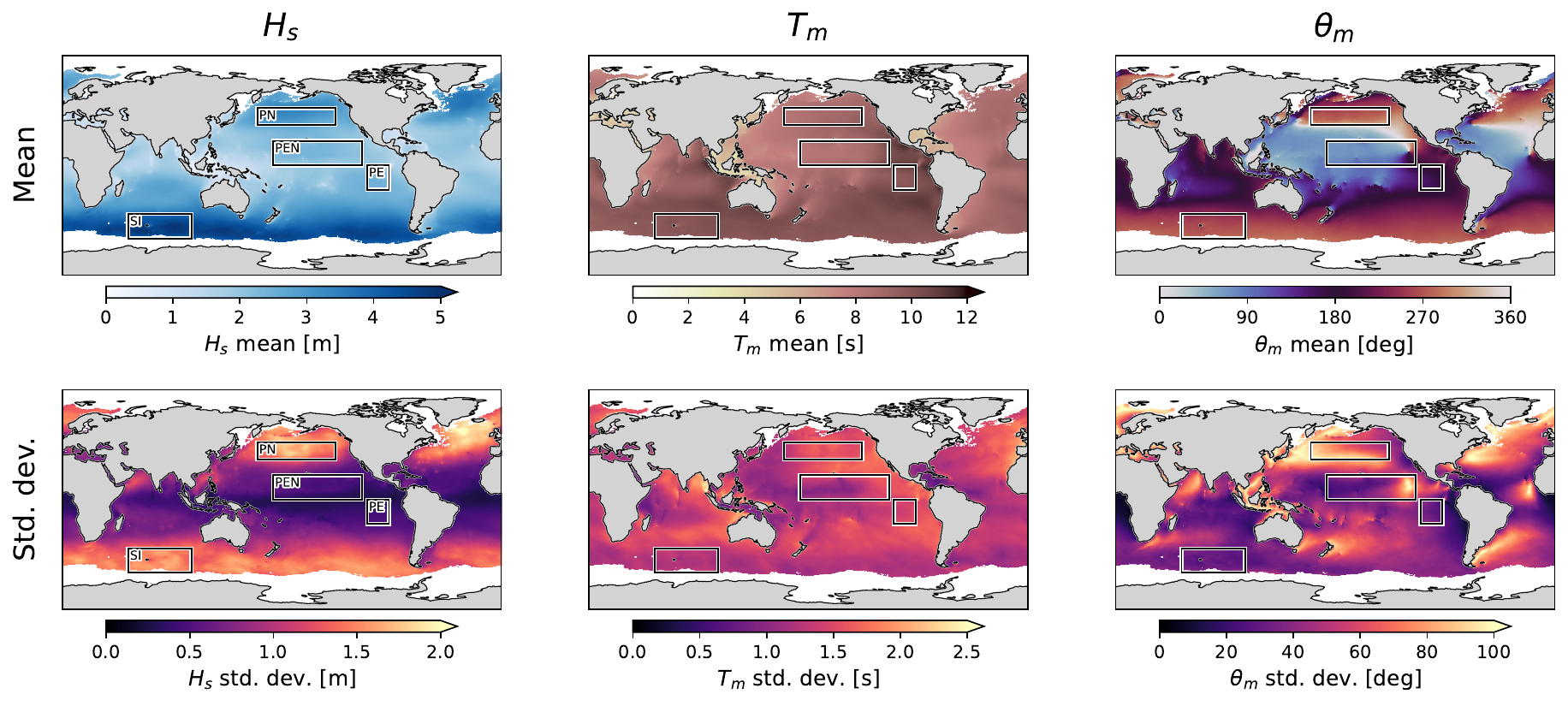}
    \caption{
    Annual mean (top row) and standard deviation (bottom row) of significant wave height $H_s$, mean wave period $T_m$, and mean wave direction $\theta_m$ in the hindcast, computed from 5-day sampling over 2020--2025. Individual tropical cyclones leave visible tracks in the standard deviation map, which reflects the limited sampling of six years at 5-day intervals. Boxes mark the regions used in Figure~\ref{fig:seasonal} (after \cite{young1999}): North Pacific (PN; 33–47°N, 160°E–136°W), northern Equatorial Pacific (PEN; 0–20°N, 173°E–114°W), Eastern Pacific (PE; 20°S–0°, 110–92°W) and the Indian sector of the Southern Ocean (SI; 40–60°S, 54–106°E).}
    \label{fig:clim-bulk-1}
\end{figure}

\subsubsection{Time-mean climatology and bias} Figure~\ref{fig:clim-bulk-1} shows the annual mean and standard deviation of the bulk variables in the hindcast. $H_s$ is largest in the Southern Ocean, where strong westerlies blow year-round over a long fetch, and is moderately large along the Northern Hemisphere storm tracks; its standard deviation is largest in the same regions. $T_m$ is largest in the Southern Ocean and in swell-dominated regions such as the eastern Pacific and the west coasts of the Americas, and its variability is largest in swell zones, where swell from remote storms arrives episodically. The mean direction $\theta_m$ divides the ocean into zones dominated by different wave systems, and its variability is largest along the boundaries between these zones.

\begin{figure}[!ht]
    \centering
    \includegraphics[width=1\linewidth]{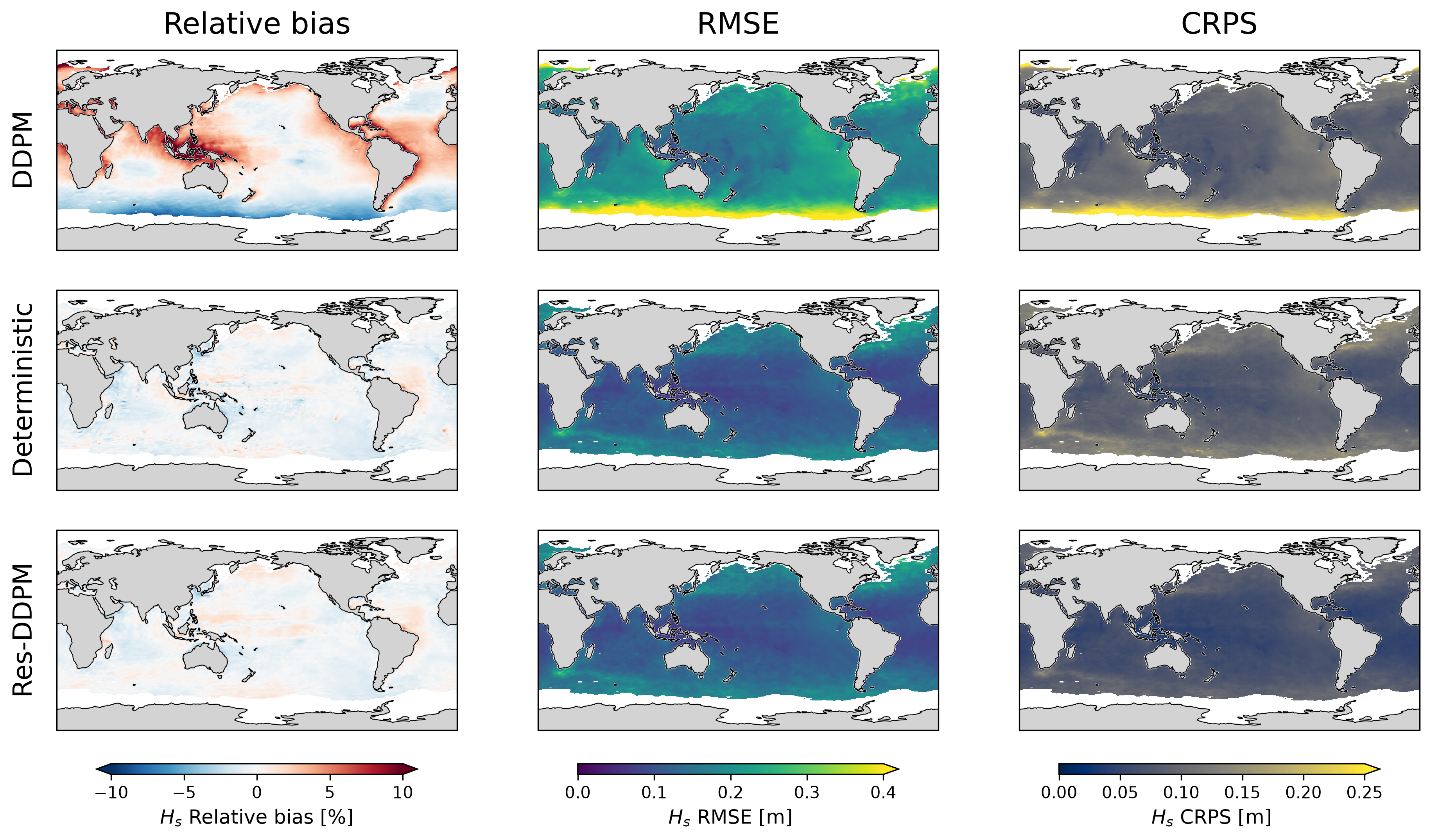}
    \caption{Time-mean skill maps of significant wave height $H_s$ for the stand-alone DDPM (top), the deterministic model (middle) and res-DDPM (bottom) over 2020--2025 (every 5 days). Left: bias of the ensemble mean (model minus truth) as a percentage of the local time-mean truth; middle: RMSE of the ensemble mean; right: CRPS (mean absolute error for the deterministic model).}
    \label{fig:skill-map-hs}
\end{figure}

Figure~\ref{fig:skill-map-hs} shows the time-mean relative bias, RMSE and CRPS of $H_s$ for the three models. The stand-alone DDPM has a systematic regional bias. $H_s$ is underestimated in the Southern Ocean, by 5--10\% and more near the sea-ice edge, and overestimated by more than 10\% in semi-enclosed and coastal seas. 
% most prominently around the Maritime Continent, in the northern Indian Ocean, the Gulf of Mexico and the Caribbean, and along the eastern boundaries of the ocean basins. 
In other words, DDPM biases regions with large $H_s$ low and regions with small $H_s$ high, pulling the time-mean field towards intermediate values. 
The bias of $T_m$ has the same spatial structure with a smaller amplitude, reaching about $-5\%$ in the Southern Ocean and more than $+5\%$ in marginal seas (Supplementary Figure~1). 
The Southern Ocean bias dominates the RMSE and CRPS maps of DDPM for $H_s$, where the RMSE exceeds 0.4~m. 
% Because positive and negative regional biases cancel, the global-mean bias of DDPM is small ($-0.2\%$ for $H_s$ and $-0.6\%$ for $T_m$; Table~\ref{tab:skill_option1}) and does not reveal this error. 
By contrast, the time-mean biases of the deterministic model and res-DDPM are within a few percent everywhere and show no coherent large-scale pattern. Since res-DDPM inherits its mean state from the deterministic prediction, removing the bias of the stand-alone DDPM is the main advantage of the residual formulation.

\begin{figure}[!ht]
    \centering
    \includegraphics[width=1\linewidth]{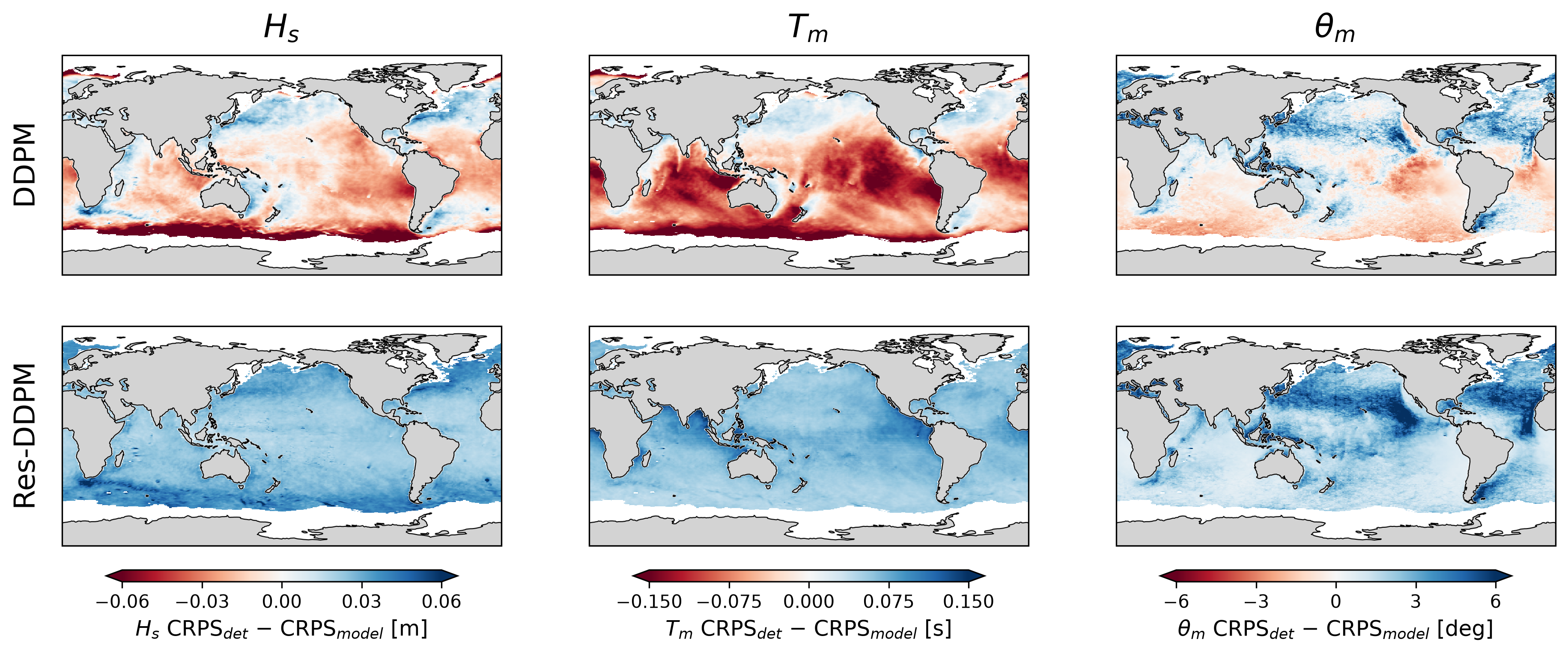}
    \caption{CRPS gain of the diffusion models relative to the deterministic model, $\mathrm{CRPS}_{\mathrm{det}} - \mathrm{CRPS}_{\mathrm{model}}$, for the bulk variables over 2020--2025 (every 5 days). Blue indicates that the diffusion model improves on the deterministic model.}
    \label{fig:crps-gain}
\end{figure}

The maps of CRPS gain (Figure~\ref{fig:crps-gain}) show where the diffusion models gain or lose skill relative to the deterministic model. Weighted by area, the stand-alone DDPM improves on the deterministic model over only 28\% ($H_s$), 19\% ($T_m$) and 51\% ($\theta_m$) of the ice-free ocean, mainly along the Northern Hemisphere storm tracks for $H_s$ and $T_m$, and is worse elsewhere. Removing this regional bias of the stand-alone DDPM will require improving its formulation, potentially through stronger conditioning (e.g., classifier-free guidance), improved samplers, or post hoc bias correction. Res-DDPM improves the CRPS everywhere for all three variables. The gain for $\theta_m$ is largest in the North Pacific, the tropical Pacific and the North Atlantic, which coincide with the regions of large directional variability in Figure~\ref{fig:clim-bulk-1}: where multiple wave systems compete, a probabilistic prediction is most beneficial. However, res-DDPM is more under-dispersed than the stand-alone DDPM, with SSR between 0.52 and 0.75 for the bulk variables, compared with 0.93 for DDPM (see Supplementary Note~1 for a discussion of the possible causes).

The CRPS gain of DDPM relative to the deterministic model differs between the bulk and the partition variables (Supplementary Tables~1 and 2). For bulk variables, the deterministic model is already a skillful predictor, and the stand-alone DDPM has a higher CRPS than the deterministic model because of its regional bias, whereas res-DDPM, which shares the mean state of the deterministic model, reduces the CRPS. For the partition variables, in contrast, even the stand-alone DDPM has a lower CRPS than the deterministic model for five of the seven variables, with the largest gains for $\theta_{p2}$ (24.1\%), $T_{p2}$ (14.2\%) and $\theta_{p1}$ (12.4\%), and res-DDPM reduces the CRPS by 24.4--34.9\%. The partitioned fields contain sharp spatial edges, such as the boundaries between crossing and non-crossing seas and between partitions of different directions, whose exact location is uncertain. The deterministic model, which approximates the conditional mean, smooths these edges, whereas single members of both the stand-alone DDPM and res-DDPM reproduce them (Supplementary Figure~2).

\begin{figure}[!ht]
    \centering
    \includegraphics[width=1\linewidth]{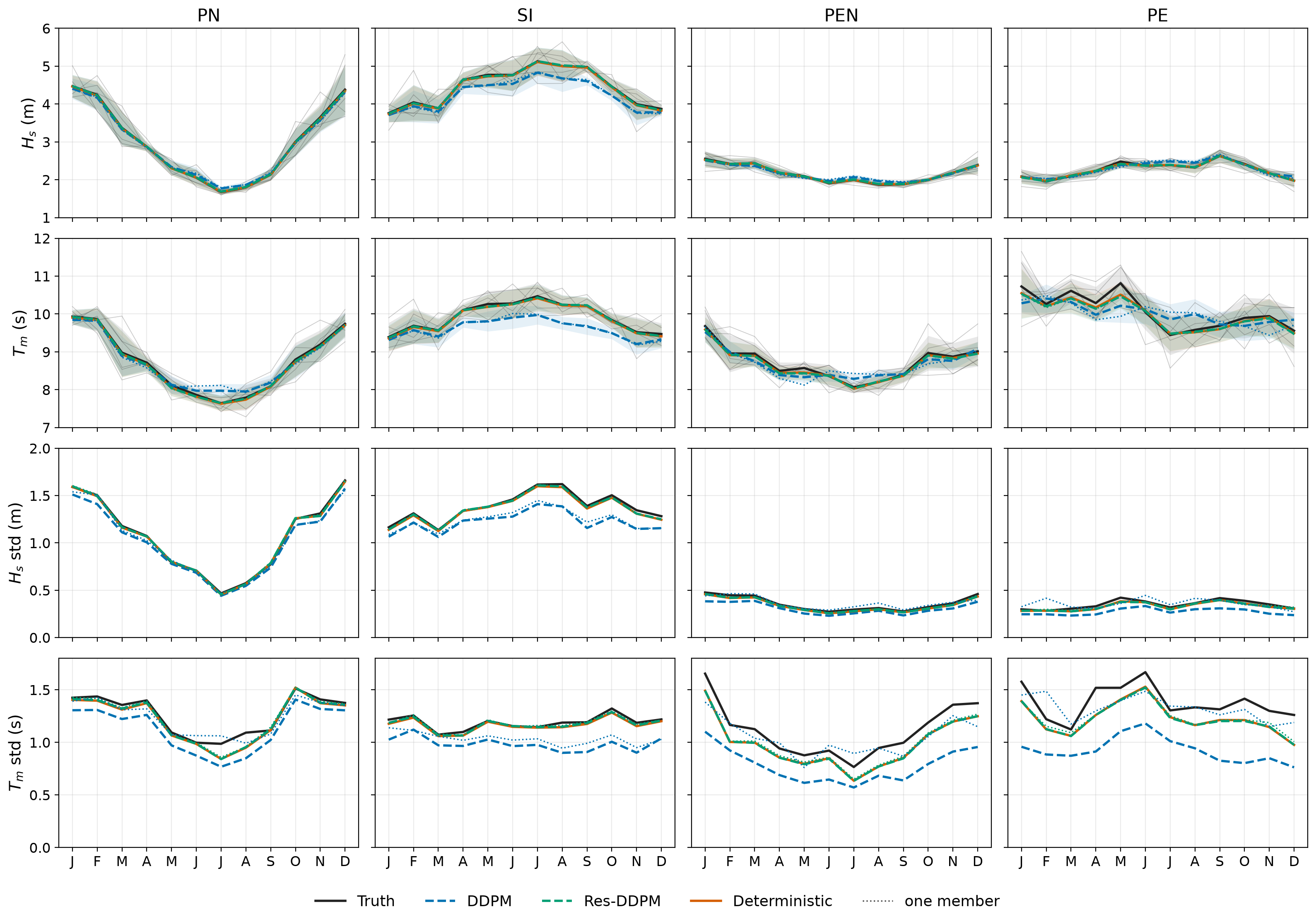}
    \caption{Seasonal cycle of the regional mean (top two rows) and regional variability (bottom two rows) of significant wave height $H_s$ (first and third rows) and mean wave period $T_m$ (second and fourth rows) in the four regions marked in Figure~\ref{fig:clim-bulk-1}. Regional means are area-weighted averages over each region, computed from samples every 5 days over 2020--2025. Shading shows $\pm 1$ interannual standard deviation for both hindcast and DDPM; thin grey lines show individual years for hindcast. Regional variability is the standard deviation over all pixels and dates within each calendar month, pooled across 2020--2025. Lines show the hindcast (black), the deterministic model (orange), the 10-member ensemble means of the stand-alone DDPM (blue dashed) and res-DDPM (green dashed), and a single ensemble member of each (dotted).}
    \label{fig:seasonal}
\end{figure}

\subsubsection{Seasonal cycle} 
To test whether the models reproduce the wave climate beyond the time mean, we compare the regional seasonal cycles in four regions of \cite{young1999} with distinct seasonal behaviour (boxes in Figure~\ref{fig:clim-bulk-1}). The deterministic model and res-DDPM follow the hindcast regional mean of $H_s$ and $T_m$ in all four regions (Figure~\ref{fig:seasonal}). The ensemble mean of res-DDPM nearly coincides with the deterministic model, as expected since the res-DDPM ensemble averages out to the deterministic prediction. The stand-alone DDPM reproduces the seasonal cycle of the regional mean $H_s$ in the North Pacific (PN), the northern equatorial Pacific (PEN) and the eastern Pacific (PE), including the boreal-winter maximum in PN. In the Indian sector of the Southern Ocean (SI), it underestimates $H_s$ throughout the year, by up to about 0.4~m in austral winter and spring, and $T_m$ by up to about 0.5~s, consistent with the time-mean bias in Figure~\ref{fig:skill-map-hs} and Supplementary Figure~1. 

The regional variability (standard deviation over all grid points and dates in each calendar month) shows a different picture. The deterministic model and the res-DDPM ensemble mean reproduce the variability of $H_s$ in all four regions and that of $T_m$ in SI and in PN outside boreal summer, where the variability is driven mainly by synoptic wind variability and a deterministic wind-to-wave mapping is sufficient. In the swell-dominated regions PEN and PE, both underestimate the variability of $T_m$. The DDPM ensemble mean underestimates the variability of $T_m$ in all four regions, most strongly in PE (by 20--45\%), and that of $H_s$ in SI by 10--15\%. A single DDPM ensemble member, however, is considerably more variable than the ensemble mean, and in PEN and PE its $T_m$ variability is close to that of the hindcast. This is where the stochastic component of the diffusion model matters: the ensemble mean, like the deterministic prediction, is a smoothed estimate of the conditional mean, while individual members retain the variability of the target.

In summary, the stand-alone DDPM produces realistic individual samples and a nearly calibrated ensemble, but it has a regional climatological bias of up to about 10\% in $H_s$: too low in the Southern Ocean and too high in marginal and coastal seas. Res-DDPM removes this bias by inheriting the mean state from the deterministic model, and its added stochastic component yields the lowest CRPS of the three models for all bulk and partition variables (Supplementary Tables~1 and 2), at the cost of an under-dispersed ensemble. We therefore suggest res-DDPM as an alternative to the stand-alone DDPM in applications where the wave climatology is important.

\section{Discussion}

In this work, we have proposed SWADDL, a diffusion model that learns to sample sea state from the complex conditional distribution given global wind history, trained on multi-decadal hindcast data. Motivated by the need to characterize multi-modal sea states and to enable sea-state-dependent parameterizations, we go beyond the bulk variables targeted by existing deep learning approaches and predict an array of variables that characterize different aspects of the full 2D wave spectrum. We find that not all sea state variables are equally difficult to predict: wave height is easier than mean wave period; bulk variables are easier than partitioned variables; the crossing sea index is predicted with skill, but the mean variables of individual partitions---particularly the less energetic one---remain challenging; and derived variables corresponding to different higher-order moments of the frequency spectrum exhibit varying levels of dependence on wind history.

% Discuss some of the modeling choice
Given that the sea state is largely dictated by wind forcing, for the temporal evolution of wave field, our approach is to condition the sampling on a rolling window of global wind history rather than adopting the autoregressive time-stepping framework used by most existing AI-based wave models (see recent review in \cite{jiang_data_2026}). This involves a trade-off: for bulk variables where autoregressive baselines exist, our RMSE is comparable at long lead times but larger at short lead times. However, foregoing autoregression prevents error accumulation and provides the flexibility to query any subset of sea state variables at any desired temporal interval. Most importantly, it enables prediction of partition-related and higher-order derived variables, which are not easily amenable to autoregressive frameworks. One thing to keep in mind is that the spectrally integrated variables predicted here do not obey a closed-form evolution equation, even though the underlying 2D spectrum evolves according to the wave action balance equation.
% conditioning directly on wind history bypasses this difficulty.

% Discuss on conditioning
For the wind history window used for conditioning, we choose 10 days of daily sampled instantaneous zonal and meridional wind fields, guided by typical swell propagation times across ocean basins. Ablation tests (Supplementary Note~2 and Supplementary Figure~3) for the bulk variables show that sampling wind history every 2 days instead of daily increases the CRPS by about 32\% for $H_s$, 21\% for $T_m$ and 9\% for $\theta_m$. We attribute this to the evolution of synoptic weather systems driving high waves, which move and intensify on time scales of about a day and are therefore poorly resolved by 2-day sampling. In contrast, shortening the wind history to 5 days have negligible effects on CRPS. A possible explanation is that swells generated by very remote storms (which we expect to depend on a longer wind history) make up only a small fraction of the total pixels and contribute little to the total score. 

By examining the climatology of samples generated over the multi-year testing window, we find that there is a regional climatological bias in $H_s$ from the stand-alone DDPM formulation (Section~\ref{subsec:climatology}), up to about 10\% too low in the Southern Ocean and too high in coastal seas. A residual diffusion model that samples departures from a deterministic prediction removes this bias and has the lowest CRPS, although its ensemble is under-dispersed. In future works, we will investigate other approaches to remove the regional bias of the stand-alone DDPM (e.g., by strengthening the conditioning, improving the sampler, or correcting the bias post hoc) and ways to make the residual DDPM better calibrated. More broadly, SWADDL can be further refined through a more systematic exploration of the design space, including model architecture (especially the use of latent space), sampling strategies, and additional conditioning fields such as ocean currents. We present these results recognizing that the reported metrics can likely be improved with such extensions.

Despite the iterative nature of diffusion models, SWADDL has achieved substantial acceleration, with over 20-time wall-clock speedup compared to spectral wave models (see Supplementary Note~3 for detailed accounting). This acceleration will facilitate incorporation of SWADDL into coupled climate models to study the effects of waves on air-sea interaction, especially in the tropical regions where swells are an important source of potential bias and additional variability \cite{shimura_long-term_2017, jeong_impact_2026}.
We can also adapt the current global framework for more localized, storm-focused sea state estimation. We envision that in such a setting, SWADDL will help improve storm forecasting through fast iterations in a coupled atmosphere-wave-ocean system, since asymmetric wave feedback plays an important role in storm development and translation.

Finally, we emphasize the novelty of the probabilistic approach to wave forecasting presented here and its potential use for data assimilation. To date, systematic assimilation of wave measurements has largely been limited to bulk parameters, mainly significant wave height. Although preliminary efforts to assimilate partition-level wave information exist \cite{hasselmann_optimal_1997, houghton_operational_2022}, their potential has not been fully realized. The ability to predict wave partitioning at low computational cost, with associated variance estimates from the diffusion model, can facilitate the development of assimilation algorithms that fully exploit the detailed wave-measurement capabilities of modern satellite platforms such as SAR \cite{collard_monitoring_2009}.
Furthermore, the model can serve as a computationally efficient forward operator mapping surface winds to the wave field, enabling ensemble wave prediction from ensemble wind forcing. This also opens the door to inverse approaches for surface wind retrieval and the assessment of wind products based on the more readily observable wave field.

% From wind and wave observations to forecast directly \cite{schultz_can_2021, allen_end--end_2025}. We currently rely on spectral wave models. Given enough data, we could also train end-to-end (directly from wind to observables) using observations. Keeping in mind that the predictions of wave partitions, even in the most state-of-the-art wave models, are not perfect \cite{cavaleri_windwave_2020}, this direct approach may give us better insight and ways to understand the origins of numerical model errors (whether it's from the wind or model physics).
% In other words, the physical system described by the reduced dimension variables may not be first-order Markovian (or do not constitute a complete set of variables to evolve the system in an auto-regressive way, or have less skills in prediction). 
% An inverse model from waves to wind.

\section{Methods}
%Summarize the main methods or models and the datasets used.
\subsection{Sea state variables} \label{subsec:variables}
The global wave data come from a 30-year hindcast performed by LOPS Ifremer \citep{rascle_global_2013,  alday_global_2021}, based on the spectral numerical wave model WAVEWATCH-III, forced by ERA5 winds, CMEMS-GLOBCURRENT currents, SSMI ice mask, and ALTIBERG iceberg mask. The resolution of all output fields is 320 in latitude (from 78$^\circ$S to 83$^\circ$N) and 720 in longitude. 

We use the following integral variables from the hindcast output as sea state prediction targets, which characterize different aspects of the sea state.

\subsubsection{Bulk (mean) variables}
The bulk (mean) variables, including significant wave height $H_s$, mean wave period $T_m$, and mean wave propagation direction $\theta_m$, are defined as integrals over the entire frequency-direction spectrum.
% Definition of mean properties as $i$-th spectral moments of partition $k$:
% \begin{equation}
%     m_{ik} = \int_{0}^{\inf} \omega^{i} S_k(\omega)d\omega.
% \end{equation}
Wave energy $e$ is defined as the spectral density $F(f,\theta)$ integrated over frequency $f$ and direction $\theta$:
\begin{equation}\label{eqn:energy}
    e = \int_f\int_{\theta} F(f,\theta)\: d\theta df.
\end{equation}
The spectral average $\bar{z}$ of any quantity $z$ is defined as a weighted integral:
\begin{equation}\label{eqn:spec_aver}
    \bar{z} = \frac{1}{e}\int_f\int_{\theta} zF(f,\theta)\: d\theta df.
\end{equation}
Significant wave height $H_s$ is defined as  
\begin{equation}
    H_s = 4e^{1/2},
\end{equation}
mean wave period $T_m$ (also referred to as the wave energy period in the literature) as
% (the energy period also used in ERA5, names $T_{0m1}$ in WW3 doc)
\begin{equation}
    T_m = \overline{f^{-1}},
\end{equation}
and mean wave direction $\theta_m$ as
\begin{equation}\label{eqn:theta_m}
    \theta_m = \tan^{-1} \left( \frac{\overline{\sin \theta}}{\overline{\cos \theta}} \right),
\end{equation}
where $\overline{f^{-1}}$, $\overline{\sin \theta}$, and ${\overline{\cos \theta}}$ are defined according to Equation \ref{eqn:spec_aver}. Note that the direction is defined following the oceanic convention, where 270$^\circ$ means waves traveling from the west and 180$^\circ$ means waves traveling from the south.

\subsubsection{Partition-based variables} \label{subsection:partition_var}
Wave partitions represent wave systems generated by different sources (local winds or remote storms) that coexist at a given location and can be diagnosed from the sub-peaks within the full spectrum. In the hindcast dataset, spectral partitioning is based on a watershed algorithm \cite{the_wavewatch_iii_development_group_ww3dg_wave_2016, hanson_pacific_2009}.

% Although not routinely used as metrics for evaluating wave models, wave-partition information provides significantly more detail than bulk parameters. \textcolor{blue}{How partition-based data assimilation can help provide better wave forecasting and identify numerical model deficiencies. How to link a partition with particle-based numerical models? Spatial and temporal coherence, see Hanson and Phillips 2001.}

The wave height and mean direction for each partition are defined similarly to Equations \ref{eqn:energy} and \ref{eqn:theta_m}, but with the integral range limited to within each partition. The peak period of each partition is determined by a three-point parabolic fit to the peak of the azimuthally integrated frequency spectrum. We keep the two most energetic partitions and rank them in descending order of energy. We find that in the hindcast dataset, two wave partitions account for at least 80\% of the total wave energy in ice-free regions about 86\% of the time. The partition variables are named as $H_{sp1}, T_{p1}, \theta_{p1}, H_{sp2}, T_{p2}, \theta_{p2}$. As shown in Figure \ref{fig:var_visual}, the partitioned mean variables can be very different from the bulk variables when the full frequency-direction spectrum is multi-modal.  
% and 72% of time for 90% energy

In addition to the mean variables of each partition, we define a binary variable called the crossing sea index (CI)
\begin{equation}\label{eqn:ci}
    \text{CI} = 
    \begin{cases} 
      1, & \text{if } |\theta_{p2}-\theta_{p1}| \geq 40 \;\text{and}\; H_{sp2}/H_{sp1} \geq 0.5 \\
      0, & \text{otherwise} 
    \end{cases}
\end{equation}
which characterizes whether two collocated wave systems with comparable energy levels propagate in distinctly different directions \cite{hanson_automated_2001, cavaleri_rogue_2012}.

\subsubsection{Other derived variables}
Additional wave-related variables are useful for various downstream applications and are defined in terms of moments of the full 2D spectrum. These include the surface Stokes drift vector $(U_{ss}, V_{ss})$ and the mean square slope (MSS). This is by no means an exhaustive list of derived variables, but they serve to demonstrate how a diffusion model can estimate them without explicitly predicting the full spectrum.
% Since Stokes drift depends on the third frequency moment of the  spectrum, it cannot be directly estimated from buoy observations, which are limited to second-order moments, and is generally estimated from the spectrum. 
% Unlike wave steepness defined by significant wave height and mean wavelength, MSS receives dominant contributions from short waves (wavelength below approximately $1$ m or wave frequency above approximately $1$ Hz). 
% A full calculation of the Stokes drift requires knowledge of the directional wave spectrum; information is available from wave modeling but not from satellite or most of the many, but still relatively few, scattered buoys. Webb and Fox-Kemper (2011) find the relation between wave spectral moments and the derived Stokes drift based on a number of empirical spectral shapes and wave models.

Both surface Stokes drift and MSS are related to higher moments of the spectrum. Surface Stokes drift (for deep water) is defined as \cite{webb_wave_2011}
\begin{equation}\label{eqn:stokes}
    (U_{ss}, V_{ss}) = 2g^{-1} \int_f\int_{\theta} (\cos\theta,\:\sin\theta)( 2\pi f )^{3} F(f,\theta)\: d\theta df,
\end{equation}
which is related to the third moment of the frequency-direction spectrum. The down-wave and cross-wave mean square slope is defined as
\begin{equation}\label{eqn:mss}
    (\text{MSS}_{d}, \text{MSS}_{c}) = \int_f\int_{\theta} (\cos^2(\theta-\theta_m),\:\sin^2(\theta-\theta_m))k^2 F(f,\theta)\: d\theta df,
\end{equation}
where wavenumber $k$ is related to frequency $f$ with the deep-water wave dispersion relation $k=(2\pi f)^2/g$, which makes $\text{MSS}$ related to the fourth moment of the spectrum. 
Note that due to the limited range of resolved frequencies in WAVEWATCHIII, the MSS is integrated with only the resolved range of frequencies \cite{rascle_global_2013}.
% Due to the limited range of resolved frequencies in WAVEWATCHIII, the MSS is extrapolated to include the contribution of an unresolved spectral tail \cite{rascle_global_2013}.  

% \begin{equation}
%     (U_{ss}, V_{ss}) = 2g^{1/2} \int_k\int_{\theta} (\cos\theta,\:\sin\theta) k^{3/2} F(k,\theta)\: d\theta dk
% \end{equation}
% Surface (for finite depth)
% \begin{equation}
%     (U_{ss}, V_{ss}) = \int_k\int_{\theta} (k\cos\theta,\:k\sin\theta) \sigma  \frac{\cosh(2kd)}{\sinh^2(kd)}F(k,\theta)\: d\theta dk
% \end{equation}
% Depth dependent 
% \begin{equation}
%     (U_{ss}, V_{ss}) = g\int_k\int_{\theta} (k\cos\theta,\:k\sin\theta) \sigma  \frac{\cosh(2k(d+h))}{\sinh^2(kd)}F(k,\theta)\: d\theta dk
% \end{equation}

\subsection{Models}\label{subsec:DDPM}
\subsubsection{Conditional denoising diffusion model}
Let $\mathbf{x}_t \in \mathbb{R}^{c_x\times N_{\mathrm{lat}}\times N_{\mathrm{lon}}}$
denote the (normalized) sea-state fields at valid time $t$, consisting of one or a
combination of the variables defined in Section \ref{subsec:variables}. The conditioning fields are
\begin{equation}
  \mathbf{f}_t = \big[\mathbf{u}_{t-k\Delta},\, \mathbf{v}_{t-k\Delta}\big]_{k=0}^{K-1}
  \oplus \big[\mathbf{A}_t,\, \mathbf{h}\big]
  \in \mathbb{R}^{c_f\times N_{\mathrm{lat}}\times N_{\mathrm{lon}}},
\end{equation}
where $\mathbf{u},\mathbf{v}$ are the zonal and meridional 10-m winds, with time interval $\Delta$ over a window of $K$ ending at $t$, $\mathbf{A}_t$ is the sea-ice fraction at time $t$, $\mathbf{h}$ is the (time-invariant) ocean depth. $N_{\mathrm{lat}}$ and $N_{\mathrm{lon}}$ are the latitudinal and longitudinal dimensions, $c_x$ and $c_f$ are the state channel and forcing channel dimensions, and $\oplus$ denotes concatenation along the channel dimension ($c_f = 2K+2$). The model learns to sample from a very high-dimensional conditional distribution
\begin{equation}
  \mathbf{x}_t^{*} \sim p(\mathbf{x}_t \mid \mathbf{f}_t),
\end{equation}
independently for each $t$: no previous sea state $\mathbf{x}_{t-\Delta}$ is used as input,
so there is no autoregressive time-stepping. The sampling is achieved using the Denoising Diffusion Probabilistic Models (DDPM) \cite{ho_denoising_2020}.

\paragraph{Conditioning.} We use direct conditioning: $\mathbf{f}_t$ is concatenated with the
noisy state along the channel dimension and passed to the denoiser at every noise level.
No classifier guidance or classifier-free guidance is used, i.e.\ the network is always
trained and evaluated with the condition present, and no guidance weight is applied.

\paragraph{Training.} Given a training pair $(\mathbf{x}_0, \mathbf{f})$ (where the subscript
$0$ now denotes the clean sample), Gaussian noise $\boldsymbol{\epsilon}\sim\mathcal{N}(\mathbf{0},\mathbf{I})$
and a noise level $\sigma$ drawn from the schedule $\{\sigma_n\}_{n=1}^{N}$, we form
$\mathbf{x}_\sigma = \mathbf{x}_0 + \sigma\boldsymbol{\epsilon}$ and train the denoiser
$\boldsymbol{\epsilon}_\phi$ with parameters $\phi$ by minimizing the squared error between $\boldsymbol{\epsilon}_{\phi}$ and $\boldsymbol{\epsilon}$ over ice-free ocean points
\begin{equation}
  \mathcal{L}(\phi) = \mathbb{E}_{\mathbf{x}_0,\mathbf{f},\sigma,\boldsymbol{\epsilon}}
  \big\| \mathbf{m}\odot\big(\boldsymbol{\epsilon}_\phi(\mathbf{x}_\sigma, \sigma;\, \mathbf{f})
  - \boldsymbol{\epsilon}\big)\big\|_2^2,
\end{equation}
where $\mathbf{m}$ is the wet mask (1 over ice-free ocean, 0 over land and sea ice) and
$\odot$ is element-wise multiplication. The expectation is taken over all training samples
and noise levels.

\paragraph{Sampling.} Starting from $\mathbf{x}_N \sim \mathcal{N}(\mathbf{0}, \sigma_N^2\mathbf{I})$,
each ensemble member is obtained by iterating, for $n = N,\dots,1$,
\begin{equation}
  \mathbf{x}_{n-1} = \mathbf{x}_n - (\sigma_n - \sigma_n')\,
  \boldsymbol{\epsilon}_\phi(\mathbf{x}_n, \sigma_n;\, \mathbf{f}_t) + \eta_n \mathbf{w}_n,
  \qquad \mathbf{w}_n\sim\mathcal{N}(\mathbf{0},\mathbf{I}),
\end{equation}
with $\sigma_n' = \sigma_{n-1}^2/\sigma_n$ and $\eta_n = (\sigma_{n-1}^2 - \sigma_n'^2)^{1/2}$, in line with the DDPM sampler \cite{permenter_interpreting_2024}.
Different ensemble members differ only in the initial noise $\mathbf{x}_N$ and the injected
noise $\{\mathbf{w}_n\}$; the condition $\mathbf{f}_t$ is identical.

\subsubsection{Deterministic model}
The deterministic model $D_\psi$, with parameters $\psi$, maps the forcing directly to the
sea state, $\mathbf{x}_{\mathrm{det}} = D_\psi(\mathbf{f}_t)$. It uses the same conditional
U-Net as the diffusion model, so the two differ only in their inputs and training objective.
The input channels that carry the noisy state in the diffusion model are set to zero, and the
noise-level input is held at a fixed value ($\sigma = 1$), so the noise embedding acts as a
learned constant bias. The forcing $\mathbf{f}_t$ is concatenated as in the diffusion model.
$D_\psi$ is trained by minimizing the squared error between prediction and target over
ice-free ocean points,
\begin{equation}
  \mathcal{L}(\psi) = \mathbb{E}_{\mathbf{x}_0,\mathbf{f}}
  \big\| \mathbf{m}\odot\big(D_\psi(\mathbf{f}) - \mathbf{x}_0\big)\big\|_2^2,
\end{equation}
using the same training data, optimizer, learning rate and EMA as the diffusion model.
Because it minimizes a squared error, $D_\psi$ approximates the conditional mean
$\mathbb{E}[\mathbf{x}_t \mid \mathbf{f}_t]$. It
serves both as a deterministic baseline for the diffusion model and as the frozen backbone
$\mathbf{x}_{\mathrm{det}}$ of the residual diffusion model described next.

\subsubsection{Residual diffusion model}
The residual diffusion model (res-DDPM) samples the departure of the sea state from the
deterministic prediction rather than the field itself. Keeping $D_\psi$ frozen, each ensemble
member is constructed as
\begin{equation}
  \mathbf{x}_t^{(i)} = \mathbf{x}_{\mathrm{det}} + \mathbf{s}\odot\mathbf{r}_t^{(i)},
  \qquad \mathbf{r}_t^{(i)} \sim p(\mathbf{r}_t \mid \mathbf{f}_t, \mathbf{x}_{\mathrm{det}}),
\end{equation}
where $\mathbf{r}_t$ is the scaled residual and $\mathbf{s}\in\mathbb{R}^{c_x}$ is a
per-channel scale, broadcast over the grid. We set $\mathbf{s}$ to the root-mean-square of
$\mathbf{x}_t - \mathbf{x}_{\mathrm{det}}$ over ice-free ocean points in 400 randomly drawn
training samples, so that $\mathbf{r}_t$ has approximately unit variance, as assumed by the
noise schedule. For directional variables, the difference is wrapped to
$[-180^\circ, 180^\circ)$. The denoiser is trained with the same objective as above, with
$\mathbf{x}_0$ replaced by $\mathbf{r}_t$ and $\mathbf{x}_{\mathrm{det}}$ concatenated to the
conditioning fields. The large-scale state, including its climatological mean, is thus
inherited from the deterministic model, while the diffusion model represents the variability
around it.

\subsubsection{Hyper-parameters}
The noise level $\sigma$ is drawn from a noise schedule $\{\sigma_t\}_{t=1}^N$ between $[\sigma_\text{min}, \sigma_\text{max}]$. The noise distribution defined by the schedule is also an important design choice for the training process \cite{karras_elucidating_2022, dieleman_noise_2024}. We use the DDPM schedule for training and the linear-log schedule for sampling, with the following parameters:
\begin{itemize}
  \item DDPM: $N = 1000$, $\beta_\text{start} = 0.0001$,  $\beta_\text{end} = 0.02$
  \item Linear-log: $N = 40$, $\sigma_\text{max} = 100$, $\sigma_\text{min} = 0.01$ 
  % \item EDM \cite{karras_elucidating_2022, price_probabilistic_2025}: $\sigma_\text{min} = 80$, $\sigma_\text{max} = 0.002$, $\rho = 7$
\end{itemize}
% During training, we randomly draw from this distribution of noise levels for each batch. During sampling, we use the maximum noise level X, the minimum noise level X, and number of step 40.
% During sampling, given a total sampling steps $N_s$, the noise levels are resampled with a trailing method \cite{permenter_interpreting_2024}.
% There are many admissible schedule
% Lin, Shanchuan, et al. "Common diffusion noise schedules and sample steps are flawed." Proceedings of the IEEE/CVF winter conference on applications of computer vision. 2024. " Trailing, proposed in DPM [7], only includes the last timestep and then selects intermediate timesteps with an even interval starting from the end. "
% DPM [7]: Cheng Lu, Yuhao Zhou, Fan Bao, Jianfei Chen, Chongxuan Li, and Jun Zhu. Dpm-solver: A fast ode solver for diffusion probabilistic model sampling in around 10 steps, 2022. 3, 4
Noise $\epsilon_\theta$ is predicted using a U-Net based architecture \cite{permenter_interpreting_2024} with around 90M parameters. The architecture provides an essentially global receptive field, enabling the model to account for non-local wind history. More specifically, it is a conditional U-Net with three encoder/decoder resolution levels ($320\to160\to80$), a base channel width of 256, and channel multipliers of (1, 2, 2). Convolutions use circular padding along the longitude axis, so that the global domain is treated as periodic in longitude and features remain continuous across the zonal boundary of the grid. All fields are resized to 320×320 for the network and mapped back for evaluation. The conditioning fields are concatenated with the noisy image as inputs to the network. Noise levels are embedded using a 1024-dimensional sinusoidal embedding, injected into each of the 2 ResNet blocks at each level. The model is wrapped with EDM-style preconditioning \cite{karras_elucidating_2022}, which applies $\sigma$-dependent input/output scaling to stabilize training across the full noise schedule.
% Self-attention is applied only at the 512-channel bottleneck ($80\times80$). 
The wet mask $\mathbf{m}$ is applied to the noise $\epsilon$, the prediction $\epsilon_\theta$, and the loss calculation.
Other training hyper-parameters include:
\begin{itemize}
  \item training batch size: 4
  \item gradient accumulation steps: 4
  \item optimizer: AdamW
  \item learning rate: $1\times 10^{-4}$
  \item total training steps: 40,000
  \item Exponential Moving Average (EMA) weight decay: 0.999
\end{itemize}

\subsection{Verification metrics}\label{sec:metrics}
We evaluate the ensemble with the root-mean-square error (RMSE), the anomaly correlation coefficient (ACC), the continuous ranked probability score (CRPS), and the spread-skill ratio (SSR). Let $x_{i,j,k}^m$ denote ensemble member $m$ ($m = 1,\dots,M$) at latitude $i$, longitude $j$, and time $k$, and let $y_{i,j,k}$ denote the verification target. The ensemble mean is $\bar{x}_{i,j,k} = \frac{1}{M}\sum_m x_{i,j,k}^m$ and the ensemble variance is $S^{2}_{i,j,k} = \frac{1}{M-1}\sum_m(x_{i,j,k}^m - \bar{x}_{i,j,k})^2$. All evaluations are restricted to ice-free ocean through a mask $b_{i,j,k}$, which equals 1 in open ocean and 0 over land or ice.

We first define three point-wise quantities: the bias-corrected squared error of the ensemble mean, the ensemble variance $S^2_{i,j,k}$, and the unbiased ensemble CRPS \cite{gneiting_strictly_2007}:
\begin{equation}\label{eqn:pointwise}
\begin{split}
    e_{i,j,k} &= (\bar{x}_{i,j,k} - y_{i,j,k})^2 - S^2_{i,j,k}/M, \\
    c_{i,j,k} &= \frac{1}{M}\sum_{m=1}^{M}\left|x^m_{i,j,k}-y_{i,j,k}\right| - \frac{1}{2M(M-1)}\sum_{m=1}^{M}\sum_{m'=1}^{M}\left|x^m_{i,j,k}-x^{m'}_{i,j,k}\right|.
\end{split}
\end{equation}
These quantities are averaged either over time at each grid point, to produce skill maps, or over space and time, to produce global scores:
\begin{equation}\label{eqn:averages}
    \langle q \rangle_{i,j} = \frac{\sum_k b_{i,j,k}\, q_{i,j,k}}{\sum_k b_{i,j,k}}, \qquad
    \langle q \rangle = \frac{\sum_{i,j,k} a_i b_{i,j,k}\, q_{i,j,k}}{\sum_{i,j,k} a_i b_{i,j,k}},
\end{equation}
where $a_i = \cos(\text{lat}_i)$ accounts for grid-cell area. Writing $\langle\cdot\rangle$ for either average, the metrics are
\begin{equation}\label{eqn:metrics}
\begin{gathered}
    \text{RMSE} = \langle e \rangle^{1/2}, \qquad
    \text{CRPS} = \langle c \rangle, \\
    \text{SSR} = \left(\frac{\langle S^2 \rangle}{\langle e \rangle}\right)^{1/2}, \qquad
    \text{ACC} = \frac{\langle \tilde{\bar{x}}\,\tilde{y} \rangle}{\langle \tilde{\bar{x}}^2 \rangle^{1/2}\langle \tilde{y}^2 \rangle^{1/2}},
\end{gathered}
\end{equation}
where $\tilde{\bar{x}}$ and $\tilde{y}$ are anomalies relative to the monthly climatology. SSR $\approx 1$ indicates a well-calibrated ensemble, while SSR $<1$ ($>1$) indicates under- (over-)dispersion when the bias is small. Lower CRPS is better, and for a single deterministic member the CRPS reduces to the mean absolute error. For the binary CI, $\langle e \rangle$ is the Brier score (BS), and skill relative to climatology is measured by the Brier skill score $\text{BSS} = 1 - \text{BS}/\text{BS}_\text{clim}$ \cite{brier_verification_1950, price_probabilistic_2025}, where $\text{BS}_\text{clim}$ uses the monthly climatological CI probability as the prediction.

For wave directions, the ensemble mean is the circular mean $\bar\theta = \operatorname{atan2}(\overline{\sin\theta}, \overline{\cos\theta})$, and all differences (errors, spread, CRPS terms, and anomalies relative to the circularly averaged monthly climatology) are signed circular differences $[(\theta_1 - \theta_2 + 180^\circ) \bmod 360^\circ] - 180^\circ$.

% \subsubsection{Sensitivity test}
% How the loss (or MSE/SSR) scales with training data size and epochs. How the metrics vary due to sampling steps or ensemble size. \textcolor{blue}{Should we tune the noise level by the SSR score?}

\backmatter

% \bmhead{Supplementary information}
% If your article has accompanying supplementary file/s please state so here. 

% \section*{Declarations}
% \begin{itemize}
% \item Funding
% \item Conflict of interest/Competing interests
% \item Ethics approval and consent to participate
% \item Consent for publication
% \item Data availability 
% \item Materials availability
% \item Code availability 
% \item Author contribution
% \end{itemize}

\bmhead{Data availability} 
The WAVEWATCH-III hindcast used for training and evaluation is made publicly available by Ifremer at \url{https://data-dataref.ifremer.fr/ww3/GLOBMULTI_ERA5_GLOBCUR_01/GLOB-30M/}. The same files provide the 10-m wind (from ERA5), the sea-ice fraction and the ocean depth used as model inputs. The trained weights of all models in this study are archived together with the code on Zenodo \cite{wu_swaddl_2026}.

\bmhead{Code availability}
The code to preprocess the hindcast data, train the diffusion, deterministic and residual diffusion models, load the trained models and generate samples, and reproduce all figures and tables in this study (SWADDL, version 1.0.0) is archived on Zenodo at \url{https://doi.org/10.5281/zenodo.24455626} \cite{wu_swaddl_2026}, under the MIT licence. The archive also includes the trained weights and configuration of every model, the training scripts, and an example Jupyter notebook that loads a trained model, prepares the wind and auxiliary forcing fields from the hindcast files, samples an ensemble of wave state variables, and evaluates it against the hindcast.

\bmhead{Acknowledgements}
JW and LZ acknowledge funding from NSF through the Learning the Earth with Artificial Intelligence and Physics (LEAP) Science and Technology Center (STC) (Award \#2019625) and funding from Schmidt Sciences through the Multiscale Machine Learning In Coupled Earth System Modeling (M$^2$LInES) project. 
This research used resources of the National Energy Research Scientific Computing Center, a DOE Office of Science User Facility supported by the Office of Science of the U.S. Department of Energy under Contract No. DE-AC02-05CH11231 using NERSC award BER-ERCAP0036067. This research was also supported in part through the NYU IT High Performance Computing resources, services, and staff expertise.
BC was supported by the ERC Synergy project 856408-STUOD, the ESA Marine Atmosphere eXtreme Satellite Synergy project (MAXSS) 4000132954/20/I-NB, and the ESA Harmony Science Data Utilisation and Impact Study for Ocean 4000135827/21/NL.

\bmhead{Author contributions}
JW: conceptualization, methodology, model training and evaluation, visualization, draft writing.
BC: model evaluation, discussion, draft review and editing.
LZ: conceptualization, discussion, funding acquisition, draft review and editing.
All authors read and approved the final manuscript.

\bmhead{Competing interests}
The authors declare no competing financial or non-financial interests.

%%===========================================================================================%%
%% If you are submitting to one of the Nature Portfolio journals, using the eJP submission   %%
%% system, please include the references within the manuscript file itself. You may do this  %%
%% by copying the reference list from your .bbl file, paste it into the main manuscript .tex %%
%% file, and delete the associated \verb+\bibliography+ commands.                            %%
%%===========================================================================================%%

\bibliography{ref_clean}% common bib file

@article{shimura_long-term_2017,
	title = {Long-term impacts of ocean wave-dependent roughness on global climate systems},
	volume = {122},
	issn = {2169-9291},
	url = {https://onlinelibrary.wiley.com/doi/abs/10.1002/2016JC012621},
	doi = {10.1002/2016JC012621},
	language = {en},
	number = {3},
	urldate = {2026-11-23},
	journal = {J. Geophys. Res. Oceans},
	author = {Shimura, Tomoya and Mori, Nobuhito and Takemi, Tetsuya and Mizuta, Ryo},
	year = {2017},
	pages = {1995--2011},
}

@article{jeong_impact_2026,
	title = {Impact of wave coupling and the {Charnock} parameter on tropical circulations in a climate model},
	volume = {64},
	issn = {1432-0894},
	url = {https://doi.org/10.1007/s00382-026-08402-x},
	doi = {10.1007/s00382-026-08402-x},
	language = {en},
	number = {11},
	urldate = {2026-11-23},
	journal = {Clim. Dyn.},
	author = {Jeong, Taek-Bum and Puthiyaveettil, Navajyoth and Kjellsson, Joakim and Bidlot, Jean-Raymond and Park, Wonsun},
	month = oct,
	year = {2026},
	pages = {455},
}

@article {jiang_data_2026,
      author = "Haoyu Jiang and Sitian Li",
      title = {Data-driven solutions to the wave action balance equation: a physics-centric taxonomy and review of {AI} wave modelling},
      journal = {Artif. Intell. Earth Syst.},
      year = "2026",
      publisher = "American Meteorological Society",
      address = "Boston MA, USA",
      doi = "10.1175/AIES-D-26-0093.1",
      pages=   "e260093",
      url = "https://journals.ametsoc.org/view/journals/aies/aop/AIES-D-26-0093.1/AIES-D-26-0093.1.xml"
}

@article{gneiting_strictly_2007,
	title = {Strictly proper scoring rules, prediction, and estimation},
	volume = {102},
	issn = {0162-1459, 1537-274X},
	doi = {10.1198/016214506000001437},
	number = {477},
	journal = {J. Am. Stat. Assoc.},
	author = {Gneiting, Tilmann and Raftery, Adrian E},
	month = mar,
	year = {2007},
	pages = {359--378},
}

@Article{	  toba_local_1972,
  title		= {Local balance in the air-sea boundary processes},
  volume	= {28},
  issn		= {1573-868X},
  doi		= {10.1007/BF02109772},
  language	= {en},
  number	= {3},
  journal	= {J. Oceanogr.},
  author	= {Toba, Y.},
  month		= jun,
  year		= {1972},
  pages		= {109--120}
}

@Article{	  hanson_pacific_2009,
  title		= {Pacific hindcast performance of three numerical wave
		  models},
  doi		= {10.1175/2009JTECHO650.1},
  journal	= {J. Atmos. Ocean. Technol.},
  author	= {Hanson, Jeffrey L. and Tracy, Barbara A. and Tolman,
		  Hendrik L. and Scott, R. Douglas},
  month		= aug,
  year		= {2009},
  volume = {26},
  pages = {1614--1633}
}

@Article{	  snodgrass_propagation_1966,
  title		= {Propagation of ocean swell across the {Pacific}},
  volume	= {259},
  issn		= {0080-4614},
  number	= {1103},
  journal	= {Phil. Trans. R. Soc. Lond. A},
  publisher	= {The Royal Society},
  author	= {Snodgrass, F. E. and Groves, G. W. and Hasselmann, K. F.
		  and Miller, G. R. and Munk, W. H. and Powers, W. H.},
  year		= {1966},
  pages		= {431--497}
}

@Article{	  kitaigorodskii_application_1962,
  title		= {Application of the theory of similarity to the analysis of
		  wind-generated wave motion as a stochastic process},
  volume	= {1},
  journal	= {Izv. Acad. Sci. USSR Geophys. Ser.},
  author	= {Kitaigorodskii, S. A.},
  year		= {1962},
  pages		= {105--117}
}

@Article{	  brier_verification_1950,
  journal	= {Mon. Weather Rev.},
  title		= {Verification of forecasts expressed in terms of
		  probability},
  issn		= {1520-0493},
  language	= {en},
  author	= {Brier, Glenn W.},
  month		= jan,
  year		= {1950},
  volume = {78},
  number = {1},
  pages = {1--3},
  doi = {10.1175/1520-0493(1950)078<0001:VOFEIT>2.0.CO;2}
}

@Article{	  dulov_fetch-_2020,
  title		= {On fetch- and duration-limited wind wave growth: data and parametric model},
  volume	= {153},
  issn		= {14635003},
  shorttitle	= {On fetch- and duration-limited wind wave growth},
  doi		= {10.1016/j.ocemod.2020.101676},
  language	= {en},
  journal	= {Ocean Model.},
  author	= {Dulov, Vladimir and Kudryavtsev, Vladimir and Skiba,
		  Ekaterina},
  month		= sep,
  year		= {2020},
  pages		= {101676}
}

@Article{	  yurovskaya_self-similar_2023,
  title		= {A self-similar description of the wave fields generated by
		  tropical cyclones},
  volume	= {183},
  issn		= {1463-5003},
  doi		= {10.1016/j.ocemod.2023.102184},
  journal	= {Ocean Model.},
  author	= {Yurovskaya, Maria and Kudryavtsev, Vladimir and Chapron,
		  Bertrand},
  month		= jun,
  year		= {2023},
  pages = {102184}
}

@Article{	  collard_monitoring_2009,
  title		= {Monitoring and analysis of ocean swell fields from space: new methods for routine observations},
  volume	= {114},
  pages		= {C07023},
  copyright	= {Copyright 2009 by the American Geophysical Union.},
  issn		= {2156-2202},
  shorttitle	= {Monitoring and analysis of ocean swell fields from space},
  doi		= {10.1029/2008JC005215},
  language	= {en},
  number	= {C7},
  journal	= {J. Geophys. Res. Oceans},
  author	= {Collard, Fabrice and Ardhuin, Fabrice and Chapron,
		  Bertrand},
  year		= {2009}
}

@Article{	  rascle_global_2013,
  series	= {Ocean {Surface} {Waves}},
  title		= {A global wave parameter database for geophysical applications. {Part} 2: model validation with improved source term parameterization},
  volume	= {70},
  issn		= {1463-5003},
  shorttitle	= {A global wave parameter database for geophysical
		  applications. {Part} 2},
  doi		= {10.1016/j.ocemod.2012.12.001},
  journal	= {Ocean Model.},
  author	= {Rascle, Nicolas and Ardhuin, Fabrice},
  month		= oct,
  year		= {2013},
  pages		= {174--188}
}

@Article{	  badulin_physical_2014,
  title		= {A physical model of sea wave period from altimeter data},
  volume	= {119},
  copyright	= {© 2014. American Geophysical Union. All Rights
		  Reserved.},
  issn		= {2169-9291},
  doi		= {10.1002/2013JC009336},
  language	= {en},
  number	= {2},
  journal	= {J. Geophys. Res. Oceans},
  author	= {Badulin, S. I.},
  year		= {2014},
  pages		= {856--869}
}

@Article{	  munk_inconvenient_2009,
  title		= {An inconvenient sea truth: spread, steepness, and skewness
		  of surface slopes},
  volume	= {1},
  issn		= {1941-1405, 1941-0611},
  shorttitle	= {An {Inconvenient} {Sea} {Truth}},
  doi		= {10.1146/annurev.marine.010908.163940},
  language	= {en},
  number	= {1},
  journal	= {Annu. Rev. Mar. Sci.},
  author	= {Munk, W.},
  month		= jan,
  year		= {2009},
  pages		= {377--415}
}

@InCollection{	  garbe_transfer_2014,
  address	= {Berlin, Heidelberg},
  title		= {Transfer across the air-sea interface},
  isbn		= {978-3-642-25643-1},
  doi		= {10.1007/978-3-642-25643-1_2},
  language	= {en},
  booktitle	= {Ocean-{Atmosphere} {Interactions} of {Gases} and
		  {Particles}},
  publisher	= {Springer},
  author	= {Garbe, Christoph S. and Rutgersson, Anna and Boutin,
		  Jacqueline and de Leeuw, Gerrit and Delille, Bruno and
		  Fairall, Christopher W. and Gruber, Nicolas and Hare,
		  Jeffrey and Ho, David T. and Johnson, Martin T. and
		  Nightingale, Philip D. and Pettersson, Heidi and Piskozub,
		  Jacek and Sahlée, Erik and Tsai, Wu-ting and Ward, Brian
		  and Woolf, David K. and Zappa, Christopher J.},
  editor	= {Liss, Peter S. and Johnson, Martin T.},
  year		= {2014},
  pages		= {55--112}
}

@Article{rozet_score-based_2023,
  title   = {Score-based data assimilation},
  author  = {Rozet, François and Louppe, Gilles},
  journal = {Adv. Neural Inf. Process. Syst.},
  volume  = {36},
  year    = {2023}
}

@Article{	  leutbecher_ensemble_2008,
  title		= {Ensemble forecasting},
  volume	= {227},
  copyright	= {https://www.elsevier.com/tdm/userlicense/1.0/},
  issn		= {00219991},
  doi		= {10.1016/j.jcp.2007.02.014},
  language	= {en},
  number	= {7},
  journal	= {J. Comput. Phys.},
  author	= {Leutbecher, M. and Palmer, T.N.},
  month		= mar,
  year		= {2008},
  pages		= {3515--3539}
}

@Article{	  lang_aifs-crps_2026,
  title		= {{AIFS}-{CRPS}: ensemble forecasting using a model trained
		  with a loss function based on the continuous ranked
		  probability score},
  volume	= {2},
  issn		= {3005-1460},
  shorttitle	= {{AIFS}-{CRPS}},
  doi		= {10.1038/s44387-026-00073-7},
  language	= {en},
  number	= {1},
  journal	= {npj Artif. Intell.},
  author	= {Lang, Simon and Alexe, Mihai and Clare, Mariana C. A. and
		  Roberts, Christopher and Adewoyin, Rilwan and Ben
		  Bouallègue, Zied and Chantry, Matthew and Dramsch, Jesper
		  and Dueben, Peter D. and Hahner, Sara and Maciel, Pedro and
		  Prieto-Nemesio, Ana and O’Brien, Cathal and Pinault,
		  Florian and Polster, Jan and Raoult, Baudouin and Tietsche,
		  Steffen and Leutbecher, Martin},
  month		= feb,
  year		= {2026},
  pages		= {18}
}

@article{kochkov_neural_2024,
	title = {Neural general circulation models for weather and climate},
	volume = {632},
	issn = {1476-4687},
	url = {https://doi.org/10.1038/s41586-024-07744-y},
	doi = {10.1038/s41586-024-07744-y},
	number = {8027},
	journal = {Nature},
	author = {Kochkov, Dmitrii and Yuval, Janni and Langmore, Ian and Norgaard, Peter and Smith, Jamie and Mooers, Griffin and Klöwer, Milan and Lottes, James and Rasp, Stephan and Düben, Peter and Hatfield, Sam and Battaglia, Peter and Sanchez-Gonzalez, Alvaro and Willson, Matthew and Brenner, Michael P. and Hoyer, Stephan},
	month = aug,
	year = {2024},
	pages = {1060--1066},
}

@Article{	  kudryavtsev_2d_2021,
  title		= {{2D} parametric model for surface wave development under varying wind field in space and time},
  volume	= {126},
  copyright	= {© 2021. American Geophysical Union. All Rights
		  Reserved.},
  issn		= {2169-9291},
  doi		= {10.1029/2020JC016915},
  language	= {en},
  number	= {4},
  journal	= {J. Geophys. Res. Oceans},
  author	= {Kudryavtsev, Vladimir and Yurovskaya, Maria and Chapron,
		  Bertrand},
  year		= {2021},
  pages		= {e2020JC016915}
}

@Article{	  cavaleri_wind_2012,
  title		= {Wind waves in the coupled climate system},
  volume	= {93},
  issn		= {0003-0007, 1520-0477},
  doi		= {10.1175/BAMS-D-11-00170.1},
  language	= {en},
  number	= {11},
  journal	= {Bull. Am. Meteorol. Soc.},
  author	= {Cavaleri, L. and Fox-Kemper, B. and Hemer, M.},
  month		= nov,
  year		= {2012},
  pages		= {1651--1661}
}

@Book{		  janssen_interaction_2004,
  title		= {The {Interaction} of {Ocean} {Waves} and {Wind}},
  isbn		= {978-0-521-46540-3 978-0-511-52501-8 978-0-521-12104-0},
  doi		= {10.1017/CBO9780511525018},
  language	= {en},
  publisher	= {Cambridge University Press},
  author	= {Janssen, P.},
  month		= oct,
  year		= {2004}
}

@Misc{		  hahner_representing_2026,
  title		= {Representing the surface ocean in {ECMWF}'s data-driven forecasting system {AIFS}},
  doi		= {10.48550/arXiv.2604.25559},
  publisher	= {arXiv},
  author	= {Hahner, Sara and Zampieri, Lorenzo and Bidlot,
		  Jean-Raymond and Browne, Philip and Chantry, Matthew and
		  Clare, Mariana C. A. and Cook, Harrison and Dueben, Peter
		  and Furner, Rachel and Keeley, Sarah and Kousal, Josh and
		  Lang, Simon and Lessig, Christian and Mertes, Gert and
		  Mogensen, Kristian and Moldovan, Gabriel and Pelletier,
		  Charles and Pinault, Florian and Nemesio, Ana Prieto and
		  Raoult, Baudouin and Sandu, Irina and Cruz, Mario Santa and
		  Schloer, Jakob and Tietsche, Steffen and Zuo, Hao},
  month		= apr,
  year		= {2026},
  howpublished = {Preprint at \url{https://arxiv.org/abs/2604.25559}}
}

@Misc{		  dieleman_noise_2024,
  title		= {Noise schedules considered harmful},
  author	= {Dieleman, Sander},
  year		= {2024},
  howpublished = {\url{https://sander.ai/2024/06/14/noise-schedules.html}}
}

@Article{permenter_interpreting_2024,
  title   = {Interpreting and improving diffusion models from an optimization perspective},
  author  = {Permenter, Frank and Yuan, Chenyang},
  journal = {Proc. Mach. Learn. Res.},
  volume  = {235},
  pages   = {40461--40483},
  year    = {2024}
}

@article{karras_elucidating_2022,
	title = {Elucidating the design space of diffusion-based generative models},
	volume = {35},
	url = {https://proceedings.neurips.cc/paper_files/paper/2022/hash/a98846e9d9cc01cfb87eb694d946ce6b-Abstract-Conference.html},
	doi = {10.52202/068431-1926},
	urldate = {2026-11-27},
	journal = {Adv. Neural Inf. Process. Syst.},
	publisher = {Curran Associates, Inc.},
	author = {Karras, Tero and Aittala, Miika and Aila, Timo and Laine, Samuli},
	year = {2022},
	pages = {26565--26577},
}

@Article{	  hasselmann_optimal_1997,
  title		= {An optimal interpolation scheme for the assimilation of
		  spectral wave data},
  volume	= {102},
  issn		= {2156-2202},
  doi		= {10.1029/96JC03453},
  language	= {en},
  number	= {C7},
  journal	= {J. Geophys. Res. Oceans},
  author	= {Hasselmann, S. and Lionello, P. and Hasselmann, K.},
  year		= {1997},
  pages		= {15823--15836}
}

@Article{	  cavaleri_rogue_2012,
  title		= {Rogue waves in crossing seas: the {Louis Majesty} accident},
  volume	= {117},
  pages		= {C00J10},
  copyright	= {Copyright 2012 by the American Geophysical Union},
  issn		= {2156-2202},
  shorttitle	= {Rogue waves in crossing seas},
  doi		= {10.1029/2012JC007923},
  language	= {en},
  number	= {C11},
  journal	= {J. Geophys. Res. Oceans},
  author	= {Cavaleri, L. and Bertotti, L. and Torrisi, L. and
		  Bitner-Gregersen, E. and Serio, M. and Onorato, M.},
  year		= {2012}
}

@Article{	  webb_wave_2011,
  title		= {Wave spectral moments and {Stokes} drift estimation},
  volume	= {40},
  issn		= {14635003},
  doi		= {10.1016/j.ocemod.2011.08.007},
  language	= {en},
  number	= {3-4},
  journal	= {Ocean Model.},
  author	= {Webb, A. and Fox-Kemper, B.},
  month		= jan,
  year		= {2011},
  pages		= {273--288}
}

@Article{	  hanson_automated_2001,
  chapter	= {Journal of Atmospheric and Oceanic Technology},
  title		= {Automated analysis of ocean surface directional wave spectra},
  volume	= {18},
  issn		= {0739-0572, 1520-0426},
  doi		= {10.1175/1520-0426(2001)018<0277:AAOOSD>2.0.CO;2},
  language	= {EN},
  number	= {2},
  journal	= {J. Atmos. Ocean. Technol.},
  publisher	= {American Meteorological Society},
  author	= {Hanson, Jeffrey L. and Phillips, Owen M.},
  month		= feb,
  year		= {2001},
  pages		= {277--293}
}

@TechReport{	  the_wavewatch_iii_development_group_ww3dg_wave_2016,
  title		= {User manual and system documentation of {WAVEWATCH III} version 5.16},
  author	= {{The WAVEWATCH III Development Group (WW3DG)}},
  institution	= {NOAA/NWS/NCEP/MMAB},
  type		= {Tech. Note},
  number	= {329},
  address	= {College Park, MD, USA},
  year		= {2016},
  url		= {https://polar.ncep.noaa.gov/waves/wavewatch/manual.v5.16.pdf}
}

@Article{	  hell_particle--cell_2025,
  title		= {A particle-in-cell wave model for efficient sea-state estimates in {Earth} system models—{PiCLES}},
  volume	= {17},
  copyright	= {© 2025 The Author(s). Journal of Advances in Modeling
		  Earth Systems published by Wiley Periodicals LLC on behalf
		  of American Geophysical Union.},
  issn		= {1942-2466},
  doi		= {10.1029/2025MS005221},
  language	= {en},
  number	= {8},
  journal	= {J. Adv. Model. Earth Syst.},
  author	= {Hell, Momme and Fox-Kemper, Baylor and Chapron, Bertrand},
  year		= {2025},
  pages		= {e2025MS005221}
}

@Article{	  wang_physics-guided_2024,
  title		= {Physics-guided deep learning for skillful wind-wave
		  modeling},
  volume	= {10},
  doi		= {10.1126/sciadv.adr3559},
  number	= {49},
  journal	= {Sci. Adv.},
  publisher	= {American Association for the Advancement of Science},
  author	= {Wang, Xinxin and Jiang, Haoyu},
  month		= dec,
  year		= {2024},
  pages		= {eadr3559}
}

@article{bodnar_foundation_2025,
	title = {A foundation model for the {Earth} system},
	volume = {641},
	issn = {1476-4687},
	url = {https://doi.org/10.1038/s41586-025-09005-y},
	doi = {10.1038/s41586-025-09005-y},
	number = {8065},
	journal = {Nature},
	author = {Bodnar, Cristian and Bruinsma, Wessel P. and Lucic, Ana and Stanley, Megan and Allen, Anna and Brandstetter, Johannes and Garvan, Patrick and Riechert, Maik and Weyn, Jonathan A. and Dong, Haiyu and Gupta, Jayesh K. and Thambiratnam, Kit and Archibald, Alexander T. and Wu, Chun-Chieh and Heider, Elizabeth and Welling, Max and Turner, Richard E. and Perdikaris, Paris},
	month = may,
	year = {2025},
	pages = {1180--1187},
}

@Article{	  cavaleri_windwave_2020,
  title		= {Wind–wave modeling: where we are, where to go},
  volume	= {8},
  issn		= {2077-1312},
  shorttitle	= {Wind–{Wave} {Modeling}},
  doi		= {10.3390/jmse8040260},
  language	= {en},
  number	= {4},
  journal	= {J. Mar. Sci. Eng.},
  author	= {Cavaleri, Luigi and Barbariol, Francesco and Benetazzo,
		  Alvise},
  month		= apr,
  year		= {2020},
  pages		= {260}
}

@Article{	  alday_global_2021,
  title		= {A global wave parameter database for geophysical applications. {Part} 3: improved forcing and spectral resolution},
  volume	= {166},
  issn		= {1463-5003},
  shorttitle	= {A global wave parameter database for geophysical
		  applications. {Part} 3},
  doi		= {10.1016/j.ocemod.2021.101848},
  journal	= {Ocean Model.},
  author	= {Alday, Matias and Accensi, Mickael and Ardhuin, Fabrice
		  and Dodet, Guillaume},
  month		= oct,
  year		= {2021},
  pages		= {101848}
}

@Article{	  price_probabilistic_2025,
  title		= {Probabilistic weather forecasting with machine learning},
  volume	= {637},
  issn		= {0028-0836, 1476-4687},
  doi		= {10.1038/s41586-024-08252-9},
  language	= {en},
  number	= {8044},
  journal	= {Nature},
  author	= {Price, Ilan and Sanchez-Gonzalez, Alvaro and Alet, Ferran
		  and Andersson, Tom R. and El-Kadi, Andrew and Masters,
		  Dominic and Ewalds, Timo and Stott, Jacklynn and Mohamed,
		  Shakir and Battaglia, Peter and Lam, Remi and Willson,
		  Matthew},
  month		= jan,
  year		= {2025},
  pages		= {84--90}
}

@Article{	  wang_data-driven_2025,
  title		= {Data-driven rolling model for global wave height},
  volume	= {18},
  issn		= {1991-959X},
  doi		= {10.5194/gmd-18-5101-2025},
  language	= {English},
  number	= {16},
  journal	= {Geosci. Model Dev.},
  publisher	= {Copernicus GmbH},
  author	= {Wang, Xinxin and Wang, Jiuke and Lu, Wenfang and Dong,
		  Changming and Qin, Hao and Jiang, Haoyu},
  month		= aug,
  year		= {2025},
  pages		= {5101--5114}
}

@Misc{		  brenowitz_climate_2025,
  title		= {Climate in a {Bottle}: towards a generative foundation model for the kilometer-scale global atmosphere},
  shorttitle	= {Climate in a {Bottle}},
  doi		= {10.48550/arXiv.2505.06474},
  language	= {en},
  publisher	= {arXiv},
  author	= {Brenowitz, Noah D. and Ge, Tao and Subramaniam, Akshay and
		  Manshausen, Peter and Gupta, Aayush and Hall, David M. and
		  Mardani, Morteza and Vahdat, Arash and Kashinath, Karthik
		  and Pritchard, Michael S.},
  month		= jul,
  year		= {2025},
  howpublished = {Preprint at \url{https://arxiv.org/abs/2505.06474}}
}

@Article{	  zhang_ocean_2025,
  title		= {Ocean wave forecasting with deep learning as alternative to conventional models},
  volume	= {17},
  issn		= {1942-2466, 1942-2466},
  doi		= {10.1029/2025MS005285},
  language	= {en},
  number	= {11},
  journal	= {J. Adv. Model. Earth Syst.},
  author	= {Zhang, Ziliang and Yu, Huaming and Ren, Danqin and Zhang,
		  Chenyu and Sun, Minghua and Qi, Xin},
  month		= nov,
  year		= {2025},
  pages		= {e2025MS005285}
}

@article{song_score-based_2020,
	title = {Score-based generative modeling through stochastic differential equations},
	url = {https://openreview.net/forum?id=PxTIG12RRHS},
	language = {en},
	urldate = {2026-11-27},
	author = {Song, Yang and Sohl-Dickstein, Jascha and Kingma, Diederik P. and Kumar, Abhishek and Ermon, Stefano and Poole, Ben},
	month = oct,
	year = {2021},
  journal = {Int. Conf. Learn. Represent.}
}

@article{ho_denoising_2020,
	title = {Denoising diffusion probabilistic models},
	volume = {33},
	url = {https://proceedings.neurips.cc/paper/2020/hash/4c5bcfec8584af0d967f1ab10179ca4b-Abstract.html},
	urldate = {2026-11-27},
	journal = {Adv. Neural Inf. Process. Syst.},
	publisher = {Curran Associates, Inc.},
	author = {Ho, Jonathan and Jain, Ajay and Abbeel, Pieter},
	year = {2020},
	pages = {6840--6851},
}

@Article{	  houghton_operational_2022,
  title		= {Operational assimilation of spectral wave data from the {Sofar Spotter} network},
  volume	= {49},
  copyright	= {© 2022 Sofar Ocean.},
  issn		= {1944-8007},
  doi		= {10.1029/2022GL098973},
  language	= {en},
  number	= {15},
  journal	= {Geophys. Res. Lett.},
  author	= {Houghton, Isabel A. and Hegermiller, Christie and
		  Teicheira, Camille and Smit, Pieter B.},
  year		= {2022},
  pages		= {e2022GL098973}
}

@article{young1999,
  author  = {Young, Ian R.},
  title   = {Seasonal variability of the global ocean wind and wave climate},
  journal = {Int. J. Climatol.},
  volume  = {19},
  number  = {9},
  pages   = {931--950},
  year    = {1999},
  doi     = {10.1002/(SICI)1097-0088(199907)19:9<931::AID-JOC412>3.0.CO;2-O}
}

@article{young2011,
  author  = {Young, Ian R. and Zieger, Stefan and Babanin, Alexander V.},
  title   = {Global trends in wind speed and wave height},
  journal = {Science},
  volume  = {332},
  number  = {6028},
  pages   = {451--455},
  year    = {2011},
  doi     = {10.1126/science.1197219}
}

@misc{wu_swaddl_2026,
  author    = {Wu, Jiarong and Chapron, Bertrand and Zanna, Laure},
  title     = {{SWADDL}: Sampling global ocean wave state from wind history with a diffusion model (code and model weights)},
  version   = {1.0.0},
  year      = {2026},
  publisher = {Zenodo},
  doi       = {10.5281/zenodo.24455626},
  url       = {https://doi.org/10.5281/zenodo.24455626}
}
%% if required, the content of .bbl file can be included here once bbl is generated
%%\input sn-article.bbl

\end{document}